\documentclass[%
 reprint,
 superscriptaddress,
 preprintnumbers,
 nofootinbib,
 longbibliography,
 amsmath,
 amssymb,
 twocolumn,
 aps, 
 showpacs,
 pre,
]{revtex4-2}
\pdfoutput=1
\usepackage{amsthm}
\usepackage{physics}
\usepackage{graphicx}
\usepackage{orcidlink}
\usepackage{booktabs}
\usepackage{listings}

\usepackage{parskip}
\usepackage[nameinlink]{cleveref}
\hypersetup{
    colorlinks,
    citecolor=black,
    filecolor=black,
    linkcolor=black,
    urlcolor=black
}
\usepackage{xcolor}

\newtheorem{thm}{Theorem}
\newtheorem{lem}{Lemma}
\newtheorem{cor}{Corollary}

\renewcommand{\arraystretch}{1.25}

\begin{document}

\title{Which Similarity-Sensitive Entropy (Sentropy)?}

\
\author{Phuc Nguyen\orcidlink{0000-0001-9993-8434}}
 \affiliation{Department of Pathology, Beth Israel Deaconess Medical Center, Boston, MA 02215}

 \author{Josiah Couch\orcidlink{0000-0002-7416-5858}}
 \affiliation{Department of Pathology, Beth Israel Deaconess Medical Center, Boston, MA 02215}

\author{Rahul Bansal\orcidlink{0009-0006-0626-7752}}
 \affiliation{Department of Pathology, Beth Israel Deaconess Medical Center, Boston, MA 02215}

\author{Alexandra Morgan\orcidlink{0000-0001-7787-0547}}
 \affiliation{Department of Pathology, Beth Israel Deaconess Medical Center, Boston, MA 02215}

\author{Chris Tam\orcidlink{0000-0002-9126-4276}}
 \affiliation{Department of Pathology, Beth Israel Deaconess Medical Center, Boston, MA 02215}

\author{Miao Li\orcidlink{0009-0004-4951-2899}}
 \affiliation{Department of Pathology, Beth Israel Deaconess Medical Center, Boston, MA 02215}

\author{Rima Arnaout\orcidlink{0000-0002-7134-0040}}
 \affiliation{Department of Medicine, UCSF}
 \affiliation{Bakar Institute for Computational Health Sciences, UCSF }
 \affiliation{Center for Intelligent Imaging, UCSF }
 
\author{Ramy Arnaout\orcidlink{0000-0001-6955-9310}}
 \email{rarnaout@bidmc.harvard.edu}
 \affiliation{Department of Pathology, Beth Israel Deaconess Medical Center, Boston, MA 02215}
 \affiliation{Division of Clinical Informatics, BIDMC}
 \affiliation{Harvard Medical School, Boston, MA 02115}

\date{\today}

\begin{abstract}
    Shannon entropy is not the only entropy that is relevant to machine-learning datasets, nor possibly even the most important one.
Traditional entropies such as Shannon entropy capture information represented by elements' frequencies but not the richer information encoded by their similarities and differences.
Capturing the latter requires similarity-sensitive entropy (``sentropy'').
Sentropy can be measured using either the recently developed Leinster-Cobbold-Reeve framework (LCR) or the newer Vendi score (VS).
This raises the practical question of which one to use: LCR or VS. 
Here we address this question theoretically and numerically, using 53 large and well-known imaging and tabular datasets. 
We find that LCR and VS values can differ by orders of magnitude and are complementary, except in limiting cases. 
We show that both LCR and VS results depend on how similarities are scaled, and introduce the notion of ``half-distance'' to parameterize this dependence. 
We prove the VS provides an upper bound on LCR for all non-negative values of the R\'enyi-Hill order parameter, as well as for negative values in the special case that the similarity matrix is full rank. 
We conclude that VS is preferable only when a dataset's elements can be usefully interpreted as linear combinations of a more fundamental set of ``ur-elements'' or when the system that the dataset describes has a quantum-mechanical character. 
In the broader case where one simply wishes to capture the rich information encoded by elements' similarities and differences as well as their frequencies, we propose that LCR should be favored; nevertheless, for certain half-distances the two methods can complement each other.
\end{abstract}

\maketitle

\section{Introduction}
\begin{figure}
    \centering
    \includegraphics[width=1.\linewidth]{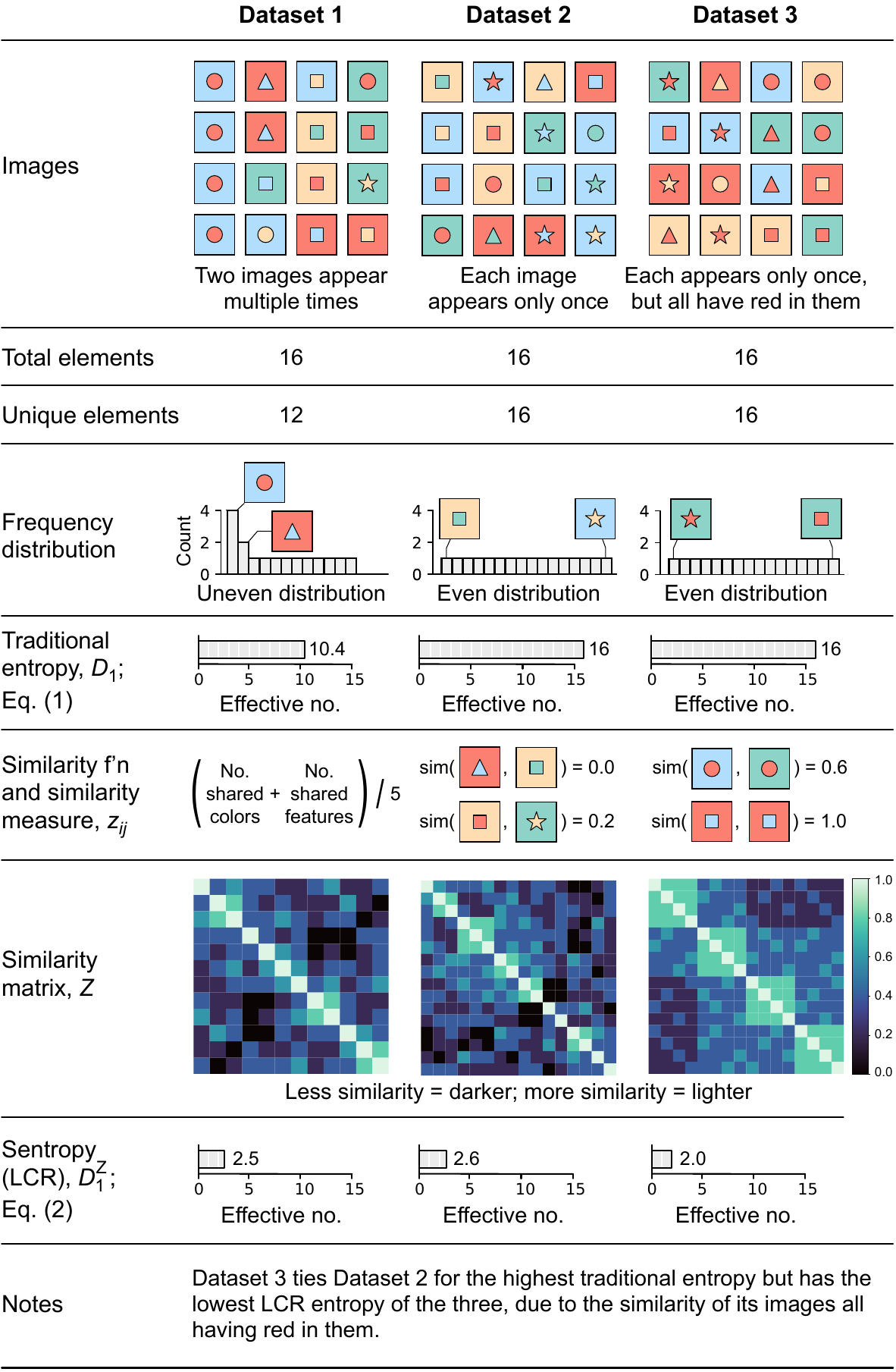}
    \caption{The concepts of element, frequency distribution, traditional entropy (here calculated at $q=1$), similarity function, similarity measure ($z_{ij}$, with examples), similarity matrix ($Z$), and sentropy (here calculated as LCR, also at $q=1$ for the most direct comparison). Entropy values are expressed in effective-number form, i.e. in units of the effective number of images present in the dataset. The choice of similarity measure is up to the investigator (see Section \ref{sec:k}); in this example, the similarity measure is defined as the normalized sum of the shared colors (0, 1, or 2) and shared features (outside color, inside color, and shape).}
    \label{fig:overview}
\end{figure}

Entropy is the foundational quantitative descriptor of information, disorder, and uncertainty in a system and finds applications across not only science and engineering but in commerce and law \cite{leinster2016maximizing, Shannon1951, Montemurro1994, tawilCompetitionCaliforniasMediCal2022}. 
Traditional entropy, as formulated by Shannon \cite{shannon1948mathematical} and generalized by Rényi \cite{renyiMeasuresEntropyInformation1961} and later Tsallis \cite{tsallis1988possible}, depends exclusively on the frequency distribution of a system’s unique elements: these entropies represent the number of bits (or nats) required to encode the shape of the distribution, with skew distributions requiring fewer bits and flatter distributions requiring more (Fig. \ref{fig:overview}).
These entropies rely solely on frequencies and therefore ignore a rich additional source of information about the system: the pairwise similarities and differences among its elements (Fig. \ref{fig:overview}).

Incorporating similarity into the calculation results in so-called similarity-sensitive entropy or sentropy. 
This was first accomplished by Leinster and Cobbold \cite{leinster2012measuring} and extended by Reeve and colleagues \cite{reeve2014partition}, in a framework we refer to as LCR \cite{leinsterEntropyDiversityAxiomatic2020}. 
By incorporating similarity, LCR can differentiate between systems that have identical frequency distributions but whose elements vary in how alike they are (Fig. \ref{fig:overview}). 
This capability is valuable in many domains \cite{nguyen2026}, especially in machine learning (ML) \cite{couch2024beyond, couch2025x} (for example, to achieve state-of-the-art performance more efficiently \cite{chinnENRICHingMedicalImaging2023}), where datasets generally consist of all-unique elements, meaning any two datasets that are the same size will have the same traditional (i.e. similarity-insensitive) entropies because their frequency distributions are both flat, even though they may differ dramatically in composition. 
Sentropy has proven to provide valuable insights into highly heterogeneous biological systems such as antibody and T-cell receptor (TCR) repertoires \cite{arnaoutFutureBloodTesting2021, aroraRepertoirescaleMeasuresAntigen2022}, in which the vast majority of elements are unique, making the distributions heavy/long-tailed or nearly flat, such that traditional entropies are less informative. 

More recently, an alternative form of sentropy to LCR has been described, consisting of the Vendi score \cite{friedman2022vendi, pasarkar2025vendiscope} and its variants or ``cousins'' \cite{pasarkar2023cousins}---hereafter collectively VS. 
This alternative raises the practical question of which form of sentropy to choose for a given system or dataset: LCR or VS. 
The present work addresses this question from several angles---conceptual, empirical, and mathematical---including by measuring LCR and VS on a wide variety of medical and non-medical imaging and tabular datasets commonly used for ML research, and comparing their values.

\section{Related Work}

Entropies fall into two classes: (\textit{i}) traditional, i.e. similarity-insensitive, and (\textit{ii}) similarity-sensitive, a.k.a. sentropy. 
Although the literature often refers to ``the'' entropy (either as a shorthand or referring specifically to Shannon entropy), both classes are actually families of measures; individual members are distinguished by how they weight frequencies and, for sentropy, by their similarity matrix as well (see Table~\ref{table:entropies}).

\subsection{Traditional, similarity-insensitive entropy}\label{sec:effective_number_form}

The best known traditional entropy is Shannon entropy: $-\sum_ip_i\log{p_i}$ \cite{shannon1948mathematical}, where $p_i$ is the frequency of unique element $i$. 
Shannon entropy can be thought of as a weighted average of elements' frequencies, with the weights being the logarithms of the frequencies themselves ($\log p_i$). 
Rényi \cite{renyiMeasuresEntropyInformation1961} generalized this into a family of entropies $H_\alpha$ of order $\alpha$---``deformations'' of Shannon entropy---in which $\alpha>1$ represents greater up-weighting of more-frequent elements relative to Shannon entropy; Shannon entropy itself is $H_{\alpha=1}$. 

Hill showed that exponentiating Rényi entropy yields the ``effective number'' of distinct elements in the dataset or system \cite{hill1973diversity}.  Effective numbers, denoted $D_q$, use a parameter $q$ (identical to Rényi’s $\alpha$) such that $q=0$ counts the distinct types (by giving zero consideration to frequencies), while larger $q$ down-weight less-frequent elements.
$D_1=\exp({H_1})$ is Shannon entropy in effective-number form.
Hill demonstrated that several other familiar statistics correspond to special cases of $D_q$ for different $q$, including Simpson’s index for $q=2$ and the Berger--Parker index for $q=\infty$ \cite{hill1973diversity}.  
For this reason, Hill's formulation is often described as a unifying framework. $D_q$ are known as the Hill numbers or D numbers (for ``diversity'') and are given by
\begin{equation}\label{Hill}
D_q(\mathbf{p}) = \exp(H_\alpha(\mathbf{p})) =
\begin{cases}
\left( \sum_{i=1}^n p_i^q \right)^{\frac{1}{1-q}}, & q \neq 1 \\[6pt]
\exp\left( -\sum_{i=1}^n p_i \ln p_i \right), & q = 1
\end{cases}
\end{equation}
where $\mathbf{p}$ denotes the frequency distribution of unique elements.  Other generalizations such as Tsallis entropy \cite{tsallis1988possible} also exist. What these traditional entropies have in common is that they are similarity-insensitive: they depend solely on $\mathbf{p}$ and ignore any relationships among the elements.

\subsection{Similarity-sensitive entropy: sentropy}

\subsubsection{The Leinster-Cobbold-Reeve framework (LCR)}

LCR \cite{leinsterEntropyDiversityAxiomatic2020} extends Hill’s framework by incorporating information about the similarities and differences of the elements within the system, which traditional entropies do not. 
A similarity matrix $Z$ is introduced, with entries $Z _{ij}\in[0,1]$ that quantify the similarity between elements $i$ and $j$.
The similarity of each element to itself is set to 1, making $Z$'s diagonal entries 1.
The resulting quantities, denoted $D_q^{Z}$, are the exponentials of similarity-sensitive versions of the Rényi entropies $H_\alpha^{Z}$ and are given by  
\begin{equation}\label{eq:LCR}
\begin{split}
D_{q}^{Z}(\mathbf{p};Z) &= \exp(H^Z_\alpha(\mathbf{p};Z)) \\
&=
\begin{cases}
\left( \sum_{i=1}^{n} p_i (Z\mathbf{p})_i^{\,q-1} \right)^{\!\frac{1}{1-q}}, & q \neq 1, \\[6pt]
\exp\!\left( -\sum_{i=1}^{n} p_i \ln (Z\mathbf{p})_i \right), & q = 1.
\end{cases}
\end{split}
\end{equation}
where $(Z\mathbf{p})_i=\sum_j z_{ij}p_j$ is the frequency-weighted average similarity of element $i$ to all the elements (including itself).  
When this average is large, element $i$ is considered ordinary, making Eq.~\ref{eq:LCR} interpretable as the average ``ordinariness'' \cite{leinster2012measuring} across all elements. 
$D_q^Z{(\mathbf{p})}$ appears widely in the recent literature, where it is variously known as: 
\begin{itemize}
    \item \textit{phylogenetic diversity} \cite{chao2010phylogenetic} in the special case where $z_{ij}$ forms an ultrametric, typically derived from a phylogenetic tree;
    \item \textit{functional diversity} \cite{chaoAttributediversityApproachFunctional2019} when the similarity pertains to elements' function, for example the binding similarity between pairs of antibodies or TCRs \cite{aroraRepertoirescaleMeasuresAntigen2022};
    \item \textit{attribute diversity} \cite{chaoAttributediversityApproachFunctional2019} when interpreting the system as a set of attribute contributions instead of frequencies; and
    \item \textit{similarity-sensitive} or \textit{similarity-aware diversity} more generally \cite{nguyen2026}, for which ``sentropy'' is a convenient shorthand.
\end{itemize}

LCR has proven useful for describing many complex systems whose empirical samples are uniform or close to uniform---i.e., where unique elements' frequency distribution is flat or nearly so---a regime where traditional entropies become less informative (Fig. \ref{fig:overview}).  
Representative applications in the life sciences include high-throughput immunology (immunomes) \cite{aroraRepertoirescaleMeasuresAntigen2022}, microbiome research (metagenomics) \cite{nguyen2026}, and medical imaging \cite{couch2024beyond}.
Specifically in ML contexts, where training sets (e.g., image collections) are often composed of unique observations (images), LCR has been shown to help identify performance predictors beyond dataset size or class balance \cite{couch2024beyond}.

\subsubsection{The Vendi score and its variants (VS)}\label{sec:VS}

VS constitutes a related but separate class of sentropic measures.  
Like LCR, VS entropies are functions of a similarity matrix, but now the matrix has dimensions $n\times n$, where $n$ is the number of observations or \textit{total} elements; for this reason we refer to it as $Z_n$ to distinguish it from the $Z$ used in LCR, which has one row/column per \textit{unique} element. (When all elements are unique, $Z=Z_n$.) We can define $Z_n$ in terms of $Z$ as
\begin{equation}\label{eq: Z_n from Z}
    \left(Z_n \right)_{i,j} = Z_{s(i), s(j)}.
\end{equation}
where $s(i)$ is the unique element of which the $i^{\text{th}}$ overall element is an instance or example.

The original VS is defined as the exponential of the Shannon entropy of the eigenvalues $\tilde{\lambda}_i = \frac{\lambda_i}{n}$ of $Z_n/n$ (where $\lambda_i$ are the eigenvalues of $Z$):
\begin{equation}\label{eq:VS}
    \mathrm{VS}{(\mathbf{p}_n;Z_n)} = \exp\!\left(-\sum_{i=1}^{n} \tilde{\lambda}_i\log \tilde{\lambda}_i\right)
\end{equation}
The division by $n$ normalizes $Z_n$ to unit trace (because as in $Z$, the diagonals of $Z_n$ equal 1).
VS is to this similarity matrix what the von Neumann entropy is to the quantum density matrix. 
Replacing the Shannon term in Eq.~\ref{eq:VS} with a Rényi entropy of order $q$ yields the so-called ``cousins'' \cite{pasarkar2023cousins} of the Vendi score:
\begin{equation}
\mathrm{VS}_{q}{(\mathbf{p}_n;Z_n)} = \left( \sum_{i=1}^{n} \tilde{\lambda}_{i}^{q} \right)^{\frac{1}{1-q}}
\end{equation}
$\mathrm{VS}_q$ is a 1-parameter family of scores with $q$ as in $D_q^Z$.

Note that LCR can also be written as a function of $Z_n$ if desired:
\begin{eqnarray}
    D_{q}^{Z}(\mathbf{p};Z) = D_{q}^{Z}(\tilde{\mathbf{p}} = \frac{1}{n};Z_n)
\end{eqnarray}
where $\tilde{\mathbf{p}}$ is the uniform distribution on the $n$ elements.

\begin{table*}[t]
\centering
\scriptsize 
\renewcommand{\arraystretch}{1.15}

\begin{tabular}{p{5.3cm} p{2.2cm} p{1.9cm} p{4.9cm}}
\toprule
\textbf{Entropy type and formula} & \textbf{Frequency weighting} & \textbf{Similarity-sensitive?} & \textbf{Notes and applications} \\
\midrule

\textbf{Shannon (Boltzmann--Gibbs)} \newline
\( H_{1}(\mathbf{p})= -\sum_{i=1}^{S} p_i \log p_i \)
& \( q=1 \) & No & Information theory, Ecology, ML loss functions, Thermodynamics \\

\textbf{Rényi entropy (\(\alpha\))} \newline
\( H_{\alpha}(\mathbf{p})=\frac{1}{1-\alpha}\log \sum_i p_i^{\alpha} \)
& Any \(q\) & No & Info-theoretic security, Ecology, Fractal analysis, Physics \\

\textbf{Tsallis entropy (\(q\)-entropy)} \newline
\( S_{q}(\mathbf{p})=\frac{1}{q-1}\big(1-\sum_i p_i^{q}\big) \)
& Any \(q\) & No & Non-extensive statistical mechanics, Turbulence, Astrophysics \\

\textbf{Quantum (von Neumann) entropy} \newline
\( S(\rho) = -\operatorname{Tr}(\rho \log \rho) \)
& \( q=1 \) & No & Quantum information, Entanglement, Quantum thermodynamics \\

\midrule

\textbf{LCR} $D_q^Z(\mathbf{p};Z)=$ \newline
$\big(\sum_i p_i(Z\mathbf{p})_i^{q-1}\big)^{1/(1-q)}$ for $q\neq1$ \newline 
$\exp(-\sum_i p_i \log(Z\mathbf{p})_i)$ for $q=1$
& Any \(q\) & Yes & Phylogenetic, attribute, and functional diversity are special cases. Biodiversity, microbiomes, language... \\

\textbf{Vendi score} \newline
\( \mathrm{VS}(\mathbf{p}_n;Z_n) = \exp{\big( -\sum_{i=1}^{n} \tilde{\lambda}_{i} \log \tilde{\lambda}_{i} \big)} \)
& \(q=1\) & Yes & \( \tilde{\lambda}_i \) are eigenvalues of $Z_n=Z/N$ (positive semi-definite). ML diversity, Generative model evaluation, Ecology \\

\textbf{Cousins of the Vendi score} \newline
\( \text{VS}_q(\mathbf{x},\mathbf{k}) = \left(\sum_i \tilde{\lambda}_i^q \right)^{\tfrac{1}{1-q}} \)
& Any \(q\) & Yes & Quantum-inspired stats, Class diversity profiling \\

\midrule

\textbf{Other variants} \newline
Burg: \( \sum_i \log p_i \) \newline
KL: \( D_{KL}(p\|q)=\sum_i p_i \log\frac{p_i}{q_i} \) \newline
CRE: \( -\int_0^\infty \bar{F}(x)\log \bar{F}(x)\,dx \)
& Partly & Partly & Spectral estimation (Burg); Bayesian inference (KL); Survival analysis (CRE) \\

\bottomrule
\end{tabular}

\caption{Summary of major entropy families, their frequency-weighting forms, similarity sensitivity, and main applications.}
\label{table:entropies}
\end{table*}

\section{Theorem: VS bounds LCR}\label{sec: math}


We now present our main theoretical result:
\begin{thm}\label{thm: Vendi bounds LCR}
    Let $Z$ be an $n \times n$ ($n>0$) positive semi-definite (PSD) similarity matrix, i.e. a PSD symmetric matrix with entries between $0$ and $1$ whose diagonal entries equal $1$. Denote LCR and VS at order $q$ as $D_q(Z, p=\frac{1}{n})$ and $\text{VS}_{q}(Z)$, respectively, on the uniform distribution. Then for $q\geq 0$ (and for $q<0$ if $Z$ is full rank):
    \begin{equation}
        VS_q(Z) \geq D_q(Z, \frac{1}{n})
    \end{equation}
\end{thm}

Following recent work \cite{aaronson2025, guevara2026, knuth2026claude, wang2026, chang2026}, large language models (LLMs; especially Claude Opus 4.6) contributed to the proofs below. All results were manually checked.


\begin{lem}\label{lem: R >= Z}
    Let $Z$ be an $n$-by-$n$ ($n>0$) non-negative symmetric matrix. Denote by $\lambda_i$ the (ordered) eigenvalues of $Z$, define $r_i = \sum_{j} Z_{ij}$, and let $R$ be the diagonal matrix defined by $R_{ii} = r_i$. Then $Z \preceq R$, i.e. $R - Z$ is PSD.
\end{lem}
\begin{proof}
    We must show that for any complex valued vector $v$, $v^\dag \left( R - Z \right) v \geq 0$. Computing, we find
    \begin{gather}
        v^\dag \left( R - Z \right) v = \sum_i r_i |v_i|^2 - \sum_{i,j} v_i^\dag Z_{ij} v_j \\
        = \sum_{i,j} Z_{ij} \left( |v_i|^2 -  v_i^\dag v_j \right)\\
        = \frac{1}{2} \sum_{i,j} \left[ Z_{ij} \left( |v_i|^2 -  v_i^\dag v_j \right) + Z_{ji} \left( |v_j|^2 -  v_j^\dag v_i \right) \right] \\
        = \frac{1}{2} \sum_{i,j} Z_{ij} |v_i - v_j|^2 \geq 0
    \end{gather}
    where we have used the symmetry of Z, and the final inequality follows since Z is non-negative, i.e. $Z_{ij}\geq 0$.
\end{proof}

\begin{lem}\label{lem: M=R^-1/2 Z R^1/2}
    Let Z be as in \Cref{thm: Vendi bounds LCR} and define $\lambda_i$, $r_i$, and $R$ w.r.t. $Z$ as in \Cref{lem: R >= Z}. Construct
    \begin{align}
        M = R^{-\frac{1}{2}} Z R^{-\frac{1}{2}}.
    \end{align}
    (Note that since each $r_i\geq 1$, $R$ is invertible). Then $M$ is a PSD matrix with the same rank as $Z$, all of whose eigenvalues fall in the range $[0,1]$.
\end{lem}
\begin{proof}
    Because multiplication by an invertible matrix preserves the rank, given that $R^{-\frac{1}{2}}$ is invertible, $\rank({M}) = \rank({Z})$.
    
    Note that $Z$, $R$, and $R^{-1/2}$ are real symmetric matrices, and hence Hermitian, and furthermore by assumption $Z$ is PSD. Consequently $M = R^{-\frac{1}{2}} Z R^{-\frac{1}{2}}$ is also PSD, as follows. Consider arbitrary complex vector $v$. Because $\left(R^{-\frac{1}{2}}\right)^\dag = R^{-\frac{1}{2}}$, it is true that
    \begin{align}
        v^\dag M v = v^\dag R^{-\frac{1}{2}} Z R^{-\frac{1}{2}} v = (R^{-\frac{1}{2}} v )^\dag Z (R^{-\frac{1}{2}} v )
    \end{align}
    But since $Z$ is PSD, we have 
    \begin{align}
        v^\dag M v = (R^{-\frac{1}{2}} v )^\dag Z (R^{-\frac{1}{2}} v ) \geq 0.
    \end{align}
    Since $v$ was arbitrary, this proves $M$ is PSD. 

    To show that the eigenvalues of $M$ are bounded above by 1, it suffices to show that $M \preceq I$, i.e. that $I - M$ is PSD. This follows because $Z \preceq R$, as proved in \Cref{lem: R >= Z}, but $I - M = R^{-\frac{1}{2}} \left(R - Z \right) R^{-\frac{1}{2}}$. We thus find that all the eigenvalues of $M$ must be in $[0,1]$.

    An alternative demonstration that the eigenvalues are bounded above by $1$ follows from the Perron-Frobenius theorem \cite{perron1907, frobenius1912}. Since $M$ is non-negative, the Perron-Frobenius theorem guarantees that $M$ has a (not necessarily unique) non-negative eigenvector, and that all non-negative eigenvectors have a single eigenvalue corresponding to the spectral radius of $M$. Such an non-negative eigenvector is provided by $R^{\frac{1}{2}} \mathbf{1}$ (with $\mathbf{1}$ being the all-ones vector), which we confirm by computing 
    \begin{align}
        M R^{\frac{1}{2}} \mathbf{1} = R^{-\frac{1}{2}} Z R^{-\frac{1}{2}} R^{\frac{1}{2}} \mathbf{1} = R^{-\frac{1}{2}} Z  \mathbf{1} = R^{-\frac{1}{2}} r  = R^{\frac{1}{2}} \mathbf{1}
    \end{align}
    Since the associated eigenvalue is $1$, and since we have already shown that $M$ is PSD, we find that all the eigenvalues of $M$ must be in $[0,1]$.
\end{proof}

\begin{lem}\label{lem: lem 1}
    Let $Z$ be as in \Cref{thm: Vendi bounds LCR}. Then for real $q\geq 1$,
    \begin{align}
        \tr \left( Z^{q} \right) \leq \sum_i \left( \sum_j Z_{ij} \right)^{q-1} = \mathbf{1}^\intercal \left(Z \mathbf{1}\right)^{q-1}.
    \end{align}
    Likewise, for real $0 < q < 1$,
    \begin{align}
        \tr \left( Z^{q} \right) \geq \mathbf{1}^\intercal \left(Z \mathbf{1}\right)^{q-1}.
    \end{align}
\end{lem}
\begin{proof}
    Let $M$ be as in \Cref{lem: M=R^-1/2 Z R^1/2}. \Cref{lem: M=R^-1/2 Z R^1/2} established that $M$ is PSD and has eigenvalues in $[0,1]$. Note that $Z = R^{\frac{1}{2}} M R^{\frac{1}{2}}$. By the Araki-Lieb-Thirring inequality \cite{Lieb:1976ny, Araki1990, audenaert2007}, for any PSD matrices $A$ and $B$, and for any $p\geq0$ and $q\geq1$, we have:
    \begin{align}
        \tr \left[ (A B A)^{pq} \right] \leq \tr \left[ (A^q B^q A^q)^{p} \right]
    \end{align}
    Substituting $A = R^{\frac{1}{2}}$, $B=M$, and $p=1$, for $q>1$:
    \begin{align}
        \tr \left[ Z^{q} \right] \leq \tr \left[ R^{\frac{q}{2}} M^q R^{\frac{q}{2}} \right] = \tr \left[ R^{q} M^q \right]
    \end{align}
    where the equality is by the cyclicity of the trace. However, because the eigenvalues of $M$ are in the range $[0,1]$ we have for $q>1$ that $M^q \preceq M$, meaning
    \begin{align}
        \tr \left[ Z^{q} \right] \leq \tr \left[ R^{q} M^q \right] \leq \tr \left[ R^{q} M \right] = \sum_i r_i^q \frac{Z_{ii}}{r_i} = \sum_i r_i^{q-1}
    \end{align}
    proving the claim for $q>1$. For $0<q<1$, the direction of the Araki-Lieb-Thirring inequality reverses, yielding 
    \begin{align}
        \tr \left[ Z^{q} \right] \geq \tr \left[ R^{q} M^q \right]
    \end{align}
    but also $M \preceq M^q$ (since $x^q \geq x$ for $0\leq x \leq 1$ and $q\leq1$) yielding 
    \begin{align}
        \tr \left[ Z^{q} \right] \geq \tr \left[ R^{q} M^q \right] \geq \tr \left[ R^{q} M \right] = \sum_i r_i^{q-1}
    \end{align}
    proving the claim for $0<q<1$.
\end{proof}

\begin{proof}[Proof of \Cref{thm: Vendi bounds LCR}]

We can now prove \Cref{thm: Vendi bounds LCR} case by case. 

\subsubsection*{Case 1: $q=+\infty$}
    The proof is an application of the Gershgorin circle inequality. First note that 
    \begin{align}
        \mathrm{VS}_{\infty}(Z) &= \frac{n}{\lambda_{\text{max}}} \\
        D_{\infty}(Z, \frac{1}{n}) &= \frac{n}{r_{\text{max}}}
    \end{align}
    where $\lambda_{\text{max}}$ is the largest eigenvalue of $Z$, i.e. $\lambda_{\text{max}} = \displaystyle\max_i(\lambda_i)$, and likewise $r_{\text{max}} = \displaystyle\max_i(r_i)$. However, by the Gershgorin circle inequality, every eigenvalue of $Z$ must fall within a Gershgorin interval, i.e. in one of the intervals $[Z_{ii} - \sum_{j\neq i} Z_{i,j}, Z_{ii} + \sum_{j\neq i} Z_{i,j}]$. As such, noting that $Z_{ii} + \sum_{j\neq i} Z_{i,j} = \sum_j Z_{i,j} = r_i$, we must have that every eigenvalue $\lambda_\mu$ of $Z$ is bounded by $\lambda_\mu \leq r_{\text{max}}$. But from \Cref{lem: lem 1}, this means that
    \begin{align}
        \mathrm{VS}_{\infty}(Z) = \frac{n}{\lambda_{\text{max}}} \geq \frac{n}{r_{\text{max}}} = D_{\infty}(Z, \frac{1}{n})
    \end{align} 

\subsubsection*{Case 2: $q>1$ and $0<q<1$}

We now compute
\begin{align}
    D_q(Z, \frac{1}{n}) &= \left( \sum_{i=1}^{n} \frac{1}{n} (\sum_j Z_{ij} \frac{1}{n})^{q-1} \right)^{\frac{1}{1-q}} \notag\\
    &= \left[ \frac{\mathbf{1}^\intercal \left(Z \mathbf{1}\right)^{q-1}}{n^q}   \right]^{\frac{1}{1-q}}
\end{align}
But since $f(x) = x^{\frac{1}{1-q}}$ is monotonically \textbf{decreasing} for $q>1$ (and $\frac{1}{n^q}$ is a positive constant), by \Cref{lem: lem 1} we get
\begin{align}
    \left[ \frac{\mathbf{1}^\intercal \left(Z \mathbf{1}\right)^{q-1}}{n^q}   \right]^{\frac{1}{1-q}} &\leq \left[ \frac{\tr \left( Z^{q} \right)}{n^q}   \right]^{\frac{1}{1-q}} \notag\\
    = \left[ \tr \left( \frac{Z}{n} \right)^q \right]^{\frac{1}{1-q}} &= VS_q(Z)
\end{align}
On the other hand, since $f(x) = x^{\frac{1}{1-q}}$ is monotonically \textbf{increasing} for $0<q<1$, but the inequality from \Cref{lem: lem 1} reverses in this case, we get the same inequality in this case.

\subsubsection*{Case 3: $q=1$}

We have proven the inequality for $q>1$ and $0<q<1$. Because both sides of the inequality are continuous functions of $q$ \cite{leinsterEntropyDiversityAxiomatic2020} \cite{pasarkar2023cousins}, the result for $q=1$ follows by continuity.

\subsubsection*{Case 4: $q=0$}

We begin by computing
\begin{align}
    D_0(Z, \frac{1}{n}) = \sum_i \frac{1}{r_i} \\
    VS_0(Z) = \rank (Z)
\end{align}

Construct $M$ as in \Cref{lem: M=R^-1/2 Z R^1/2}. The entries of $M$ are given by 
\begin{align}
    M_{ij} = \frac{Z_{ij}}{\sqrt{r_i r_j}},
\end{align}
with diagonal entries
\begin{align}
    M_{ii} = \frac{Z_{ii}}{r_i} = \frac{1}{r_i}.
\end{align}
Consequently, 
\begin{align}
    D_0(Z, \frac{1}{n}) = \tr (M)
\end{align}
However, \Cref{lem: M=R^-1/2 Z R^1/2} established that the eigenvalues of $M$ lie in the $[0,1]$ and that $M$ shares the same rank with $Z$. Letting $k= \rank (Z)$ and $\mu_i$ be the eigenvalues of $M$ (in descending order), we have 
\begin{align}
    \tr (M) = \sum_{i=1}^k \mu_i \leq \sum_{i=1}^k 1 = \rank (Z).
\end{align} 

proving \Cref{thm: Vendi bounds LCR} for $q=0$, completing the proof for all $q \geq 0$.

\subsubsection*{Case 5: $q\in[-\infty, 0)$ for $Z$ full rank}
    By definition, $VS_0(Z) = \rank(Z) = n$. R\'enyi entropies are known to decrease monotonically with $q$ \cite{leinsterEntropyDiversityAxiomatic2020}. Generalized VS is simply the R\'enyi entropy of the eigenvalues of $Z/n$. As such we have for $q\leq0$
    \begin{align}
        VS_q(Z) \geq VS_0(Z) = n
    \end{align}

    Finally, for $q=-\infty$, \cite{leinsterEntropyDiversityAxiomatic2020} proves that the LCR diversities likewise decrease monotonically in $q$, implying
    \begin{align}
         D_{-\infty}(Z, \frac{1}{n}) \geq D_q(Z, \frac{1}{n})
    \end{align}
    Define $r_i = \sum_j Z_{ij}$. Then $r_i \geq 1$, since the entries of $Z$ are non-negative, and the diagonal entries (one of which contributes to $r_i$) are all $1$. Thus
    \begin{align}
         D_{-\infty}(Z, \frac{1}{n}) = \frac{n}{\displaystyle\min_i r_i} \leq n
    \end{align}
    thereby completing the proof.
    
\end{proof}

\section{Experimental tests}

\subsection{Datasets and definitions}
The following imaging and tabular datasets were used for numerical experiments comparing LCR and VS:

\begin{itemize}
    \item \textbf{Image datasets.} 22 large ($\geq10,000$-image) medical and non-medical vision benchmarks were used: MNIST \cite{mnist}, Fashion-MNIST \cite{fashionmnist}, CIFAR-10 and CIFAR-100 \cite{krizhevsky2009learning} (from the \texttt{torchvision} \cite{torchvision} Python package); BloodMNIST, ChestMNIST, OctMNIST, OrganAMNIST, OrganCMNIST, OrganSMNIST, PathMNIST, and TissueMNIST from the MedMNIST collection \cite{yangMedMNISTClassificationDecathlon2021a, medmnistv2}; the Amphibia, Insecta, Mammalia, Plantae, and Reptilia subsets from iNaturalist \cite{visipediaInat_comp2017READMEmd}; the computed tomography (CT) and ultrasound (US) subsets from RadImageNet \cite{mei2022radimagenet}, the NIH ChestXRay dataset \cite{wangChestXRay8HospitalScaleChest2017}; COCO (Common Objects in Context) \cite{linMicrosoftCOCOCommon2015}; and MIDRC COVIDx CXR-4 \cite{wuCOVIDxCXR4Expanded2023}. For computational convenience while still preserving content, images in the iNaturalist collection as well as the MIDRC COVIDx CXR-4 dataset and the NIH ChestXRay dataset were downscaled to $250\times250$ pixels, and images in the COCO dataset were scaled to $100\times100$ pixels; also for convenience, for each dataset, 10,000 images were selected uniformly at random for LCR and VS calculations.
    \item \textbf{Tabular datasets.} After selecting for a maximum of 30 columns and 200,000 rows, the 31 most frequently downloaded datasets from the University of California Irvine Machine Learning Repository (UCIML)  were retrieved \cite{lucieImprovedPython2024}. For each dataset, identifier columns and any features containing non-numeric strings or time-series data were removed.
\end{itemize}

The terms ``system'' and ``dataset'' are used interchangeably (Fig. \ref{fig:overview}). The elements of image and tabular datasets are single images and single rows, respectively. Each element appears only once in these datasets; i.e. each element is also a unique element.

\subsection{Similarity matrices}

Similarity between each two (unique) elements $\textbf{x}_i$ and $\textbf{x}_j$ was calculated from the Euclidean distance (L2 norm) as follows:
\begin{equation}\label{eq:ED}
z_{ij} = e^{-k ||\textbf{x}_{i} - \textbf{x}_{j} ||_{2}}
\end{equation}
where $k$ is a parameter that was varied during experiments to test for effects on results. This similarity measure was chosen for its simplicity and for being provably positive-definite (PSD), guaranteeing valid similarity matrices for VS (see Section \ref{sec:conclusion}); LCR has no such restriction \cite{leinsterEntropyDiversityAxiomatic2020}. Imaging datasets were first embedded into a two- or three-dimensional space using uniform manifold approximation and projection (UMAP) \cite{mcinnesUMAPUniformManifold2020} and the default $k$ taken to be $2^{-1/2}$ (varying random seeds to verify robustness), making the norm in Eq. \ref{eq:ED} equivalent to the root-mean-square distance (RMSD). UMAP embeddings avoid the high dimensionality of raw pixel values, which has adverse effects on distance-based algorithms \cite{aggarwal2001surprising}. Note that UMAP projections are most often chosen to be in two dimensions. Empirically, UMAP embeddings generally place similar images near each other and different images further from each other (e.g. Figure \ref{fig:mnist}), rationalizing the use of these embeddings in the literature and validating their use here. We present 2D UMAP results in the main text and 3D in the appendix \ref{app:randomseeds}. 

Unlike images, the dimensionality of the tabular datasets used here is low, rationalizing direct Euclidean distance. The default $k$ for tabular datasets was $1$.

\subsection{Entropy and sentropy calculations}
The \texttt{sentropy} Python package \cite{nguyen2026} was used to compute LCR as the $\gamma$ diversity; this is the effective-number form of the corresponding entropy (see Section \ref{sec:effective_number_form}). The package accepts a user-provided similarity matrix and returns results for any $q$; we primarily report results for $q=1$ (Shannon-type LCR) for direct comparison with the default VS (see Section \ref{sec:VS}). VS was calculated using the \texttt{vendi-score} package \cite{friedman2022vendi, pasarkar2023cousins}. This implementation computes the eigenvalues of the similarity matrix, retains the positive spectrum, and evaluates the exponential of the Shannon entropy of those eigenvalues, yielding a scalar appropriate for comparison to LCR. 
Similarity matrices were calculated on a GPU-enabled workstation equipped with multi-core CPUs. LCR and VS computations were performed on a 2024 M4 Apple Mac mini with 64GB unified memory.

\subsection{Clustering}
HDBSCAN (Hierarchical Density-Based Spatial Clustering of Applications with Noise) \cite{mcinnesHdbscanHierarchicalDensity2017} was used to obtain a non-entropic estimate of the effective number of unique elements as the number of clusters present in each dataset. HDBSCAN adapts to varying densities and does not require \textit{a priori} specification of the cluster count, making it well-suited for heterogeneous data.



\subsection{Speed tests}
As a practical comparison, empirical run times for \texttt{sentropy} and \texttt{vendi-score} were calculated using the Python \texttt{timeit} package with 5-fold replications and measured on a 2024 M4 Apple Mac Mini with 64GB unified memory.

\subsection{LCR and VS of imaging and tabular datasets}

We first compared LCR to VS, both at $q=1$, on a simple, well-known imaging dataset: MNIST handwritten digits.
The results below are for a random 10,000-image subset, taken for computational convenience.
Both quantities were expressed in their effective-number forms to enable comparison with an independent benchmark: the number of clusters obtained by the state-of-the-art clustering algorithm HDBSCAN. 

\begin{figure}[b]
    \centering
    \includegraphics[width=1.\linewidth]{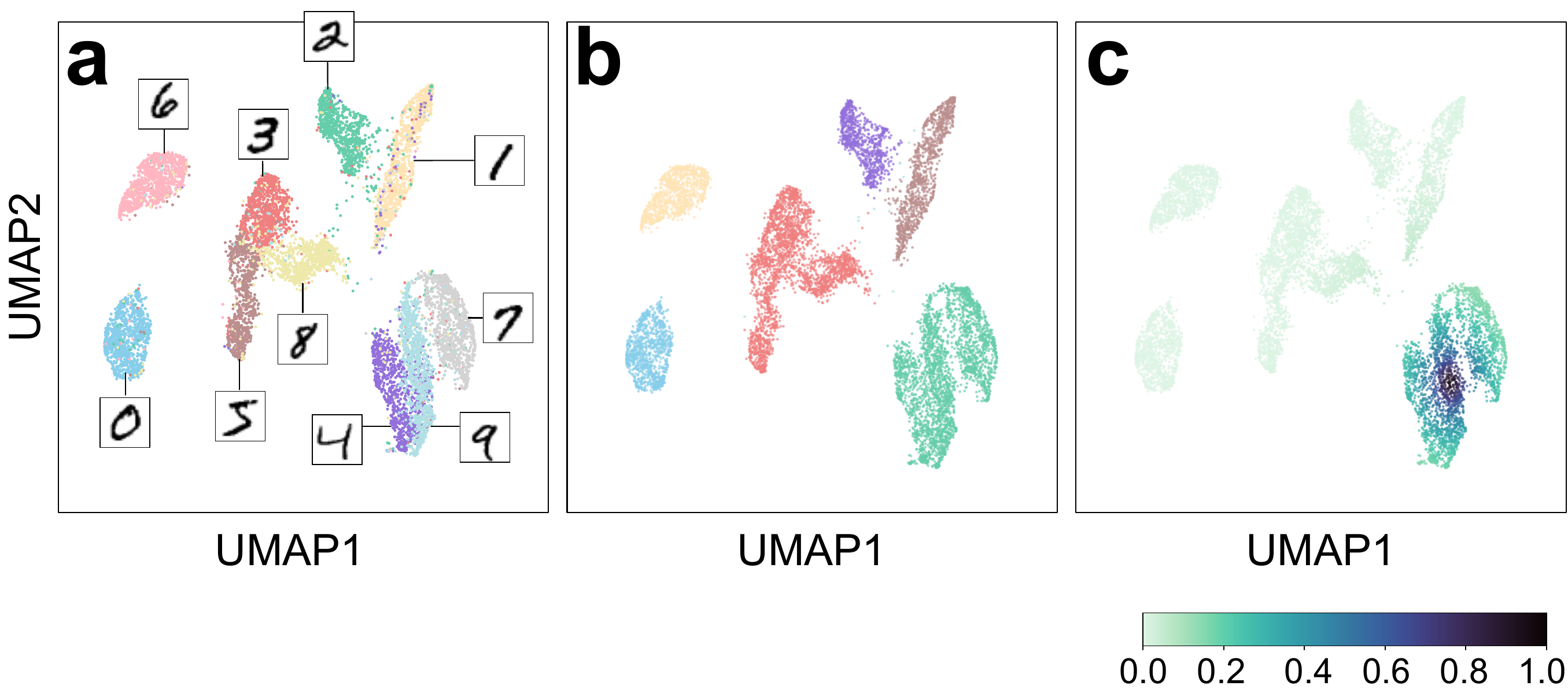}
    \caption{The first 10,000 images of the MNIST digits dataset colored (a) by digit (labeled using representative images), (b) by HDBSCAN cluster, and (c) by similarity to one of the images (a ``9''). The effective number of images in this dataset is 12.5 by LCR and 95.9 by VS.}
    \label{fig:mnist}
\end{figure}

Fig. \ref{fig:mnist}a displays a UMAP embedding of the data, coloring each digit by its label. HDBSCAN detects six clusters, successfully grouping together digits that appear visually similar (Fig. \ref{fig:mnist}b). 
LCR yields the effective number of unique images after accounting for pairwise image similarity; it allows visually similar images to each contribute less to the entropy than if they were completely distinct from each other (Fig. \ref{fig:mnist}c). 
Consequently, points that lie close together count less toward the total, whereas points separated by some distance contribute more independently, even for points that happen to be in the same cluster. 
Taken together, these considerations suggest that LCR should approximate the number of clusters---potentially exceeding it slightly when clusters are somewhat loose. 
Indeed, the similarities among the images in this 10,000-image subset result in it having an effective number of just 12.5 images according to LCR: roughly twice the number of clusters and 25\% greater than the number of classes (10). 
Mathematically, this indicates that this MNIST subset possesses the same entropy as a hypothetical dataset containing 12-13 completely dissimilar elements (i.e. with zero pairwise similarity). 
This result can be rationalized as follows: the many stylistic variations of certain digits---e.g., the different ways to write a 1, 4, or 7---increase the sentropy, while visual similarities among digits---such as the similarity between many 4s, 7s, and 9s, which cluster together per HDBSCAN---decrease it relative to a scenario in which all 10 digit classes were entirely distinct.

\begin{figure}
    \centering
    \includegraphics[width=0.8\linewidth]{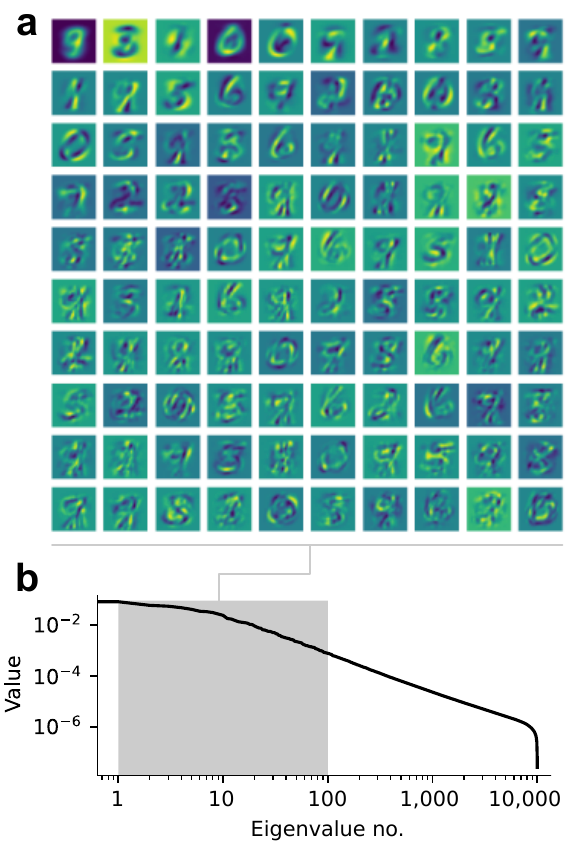}
    \caption{(a) Top 100 eigenimages for the MNIST digits dataset and (b) their eigenvalues (in the gray region), alongside the full eigenvalue spectrum. Note the log axes.}
    \label{fig:eigenimages} 
\end{figure}

In contrast, VS for this dataset was substantially higher, yielding an effective number of 95.9 elements. 
However, its interpretation differs from that of LCR: the VS value indicates that the images possess the same \textit{traditional} entropy as a collection of roughly 96 mutually orthogonal eigen- or ``ur''-images, each actual image in the dataset being a linear combination of these. 
Fig. \ref{fig:eigenimages}a displays the 100 eigen-images with the largest eigenvalues, which dominate the total entropy (Fig. \ref{fig:eigenimages}b, gray). 
Some of these eigen-images resemble recognizable digits, whereas others resemble superpositions of multiple digits that do not otherwise correspond to anything obviously familiar. 
VS is precisely the entropy of the eigenvalue spectrum, which can be interpreted as the frequency distribution of a hypothetical dataset composed of eigen-images at frequencies given by the eigenvalues. 
The total number of eigen-images equals the total number of images in the subset (10,000); the fact that the VS effective number is much smaller than this implies that only a handful of eigenvalues are large, while most are small, as confirmed by Fig. 3b.

\begin{figure}
    \centering
    \includegraphics[width=0.945\linewidth]{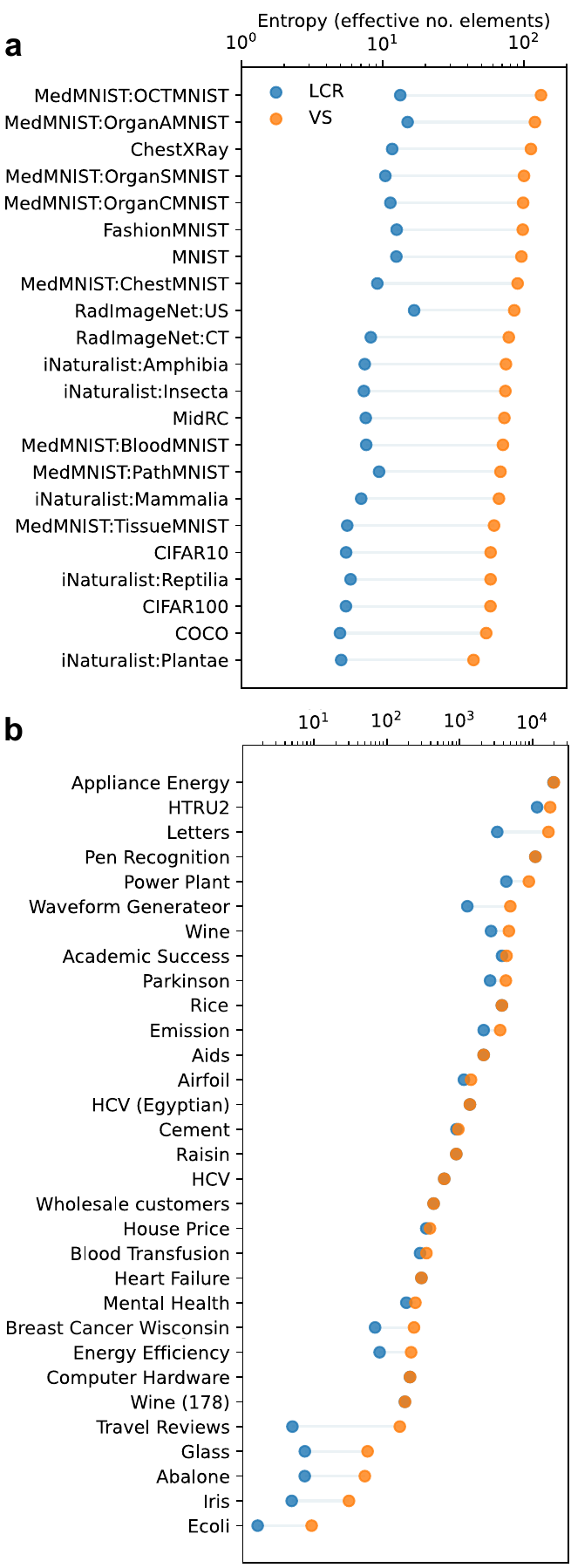}
    \caption{LCR and VS for (a) imaging and (b) tabular datasets, sorted by VS ($q=1$; default values of $k$ for each dataset type).}
    \label{fig:table}
\end{figure}

Fig. \ref{fig:table} shows LCR and VS values for 53 well known machine-learning datasets, including medical and non-medical imaging and tabular datasets, demonstrating large differences for imaging datasets (embedded in 2D UMAP projections) and smaller (and sometimes no) differences for tabular datasets (in their native space).

\subsection{Trends with LCR and VS}
\begin{figure}[t]
    \centering
    \includegraphics[width=1.\linewidth]{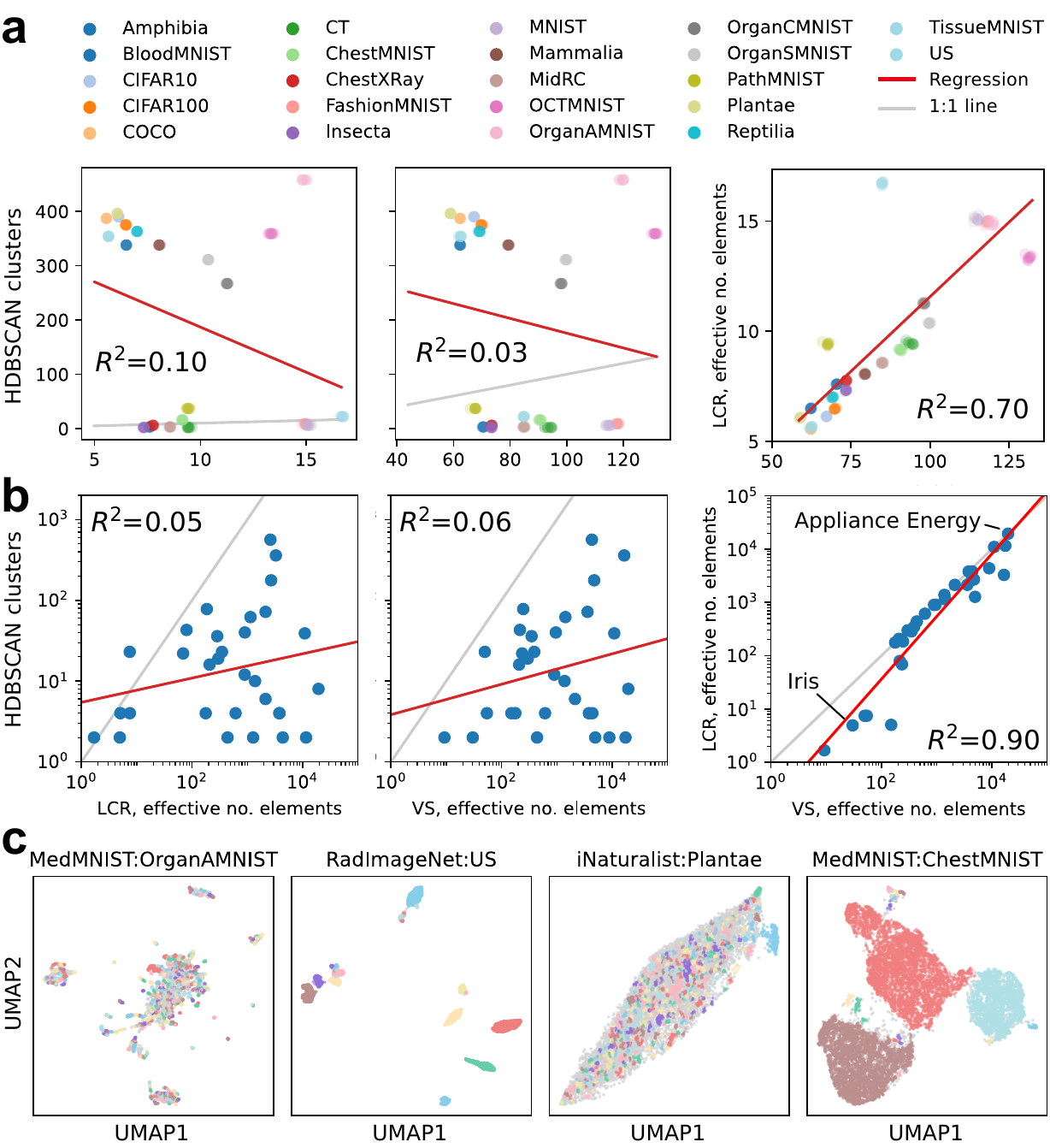}
    \caption{Correlations among LCR, VS, and HDBSCAN for (a) imaging (2D UMAP; 10 different random seeds) and (b) tabular datasets. Each point represents a dataset. All entropies are effective-number forms at $q=1$. Red line = linear regression fit; gray line = 1:1 (off the scale to the left for (a), far right). Select datasets are labeled. (c) UMAPs for labeled imaging datasets, colored by cluster. Gray points = unclustered elements.}
    \label{fig:trends}
\end{figure}

\begin{figure}[t]
    \centering
    \includegraphics[width=1.\linewidth]{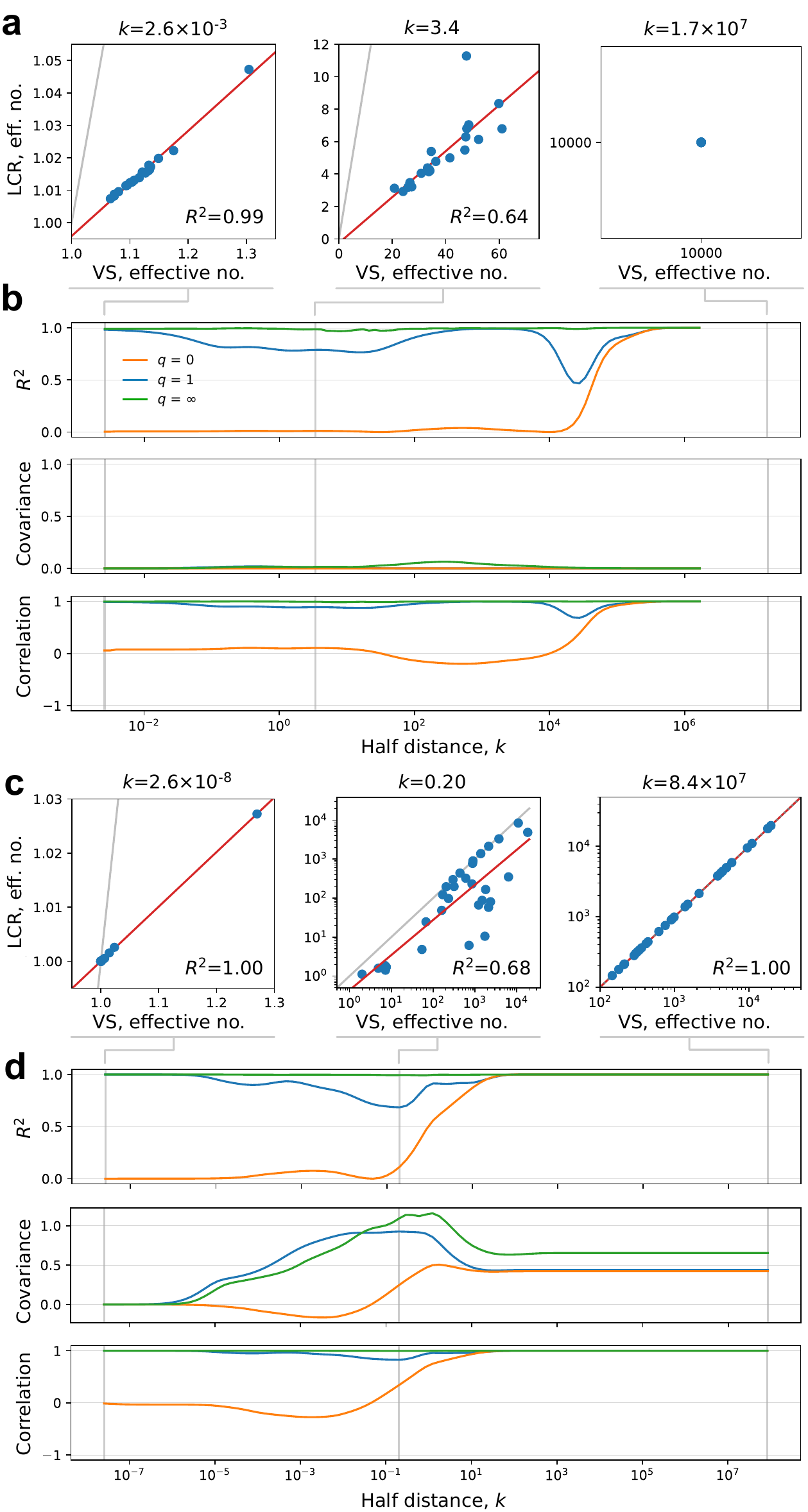}

    \caption{LCR vs. VS at a very small half-distance, an intermediate half-distance with low correlation, and large half-distance for $q=0, 1,$ and $\infty$ for (a-b) imaging and (c-d) tabular datasets. The scatterplots in (a) and (c) are for $q=1$. Red line = linear regression fit; gray line = 1:1. Each point represents a dataset. (b) and (d): vertical gray lines = the $k$ values plotted in (a) and (c).}

    \label{fig:other_hds}
\end{figure}

\begin{figure}
    \centering
    \includegraphics[width=0.9\linewidth]{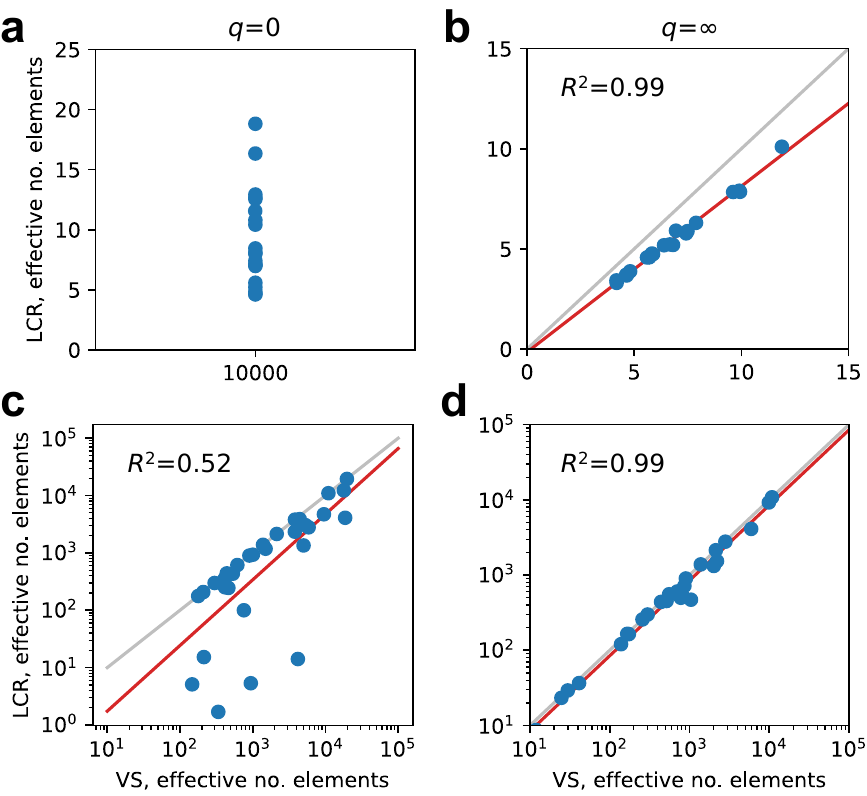}
    \caption{LCR vs. VS for $q=0$ and $\infty$ for (a-b) imaging and (c-d) tabular datasets at their default $k$ values ($1$ and $\approx2^{-1/2}$, respectively). Each point represents a dataset.}
    \label{fig:different_qs}
\end{figure}

Fig. \ref{fig:trends} presents comparisons among LCR, VS, and the number of HDBSCAN clusters across all datasets (for default $k$ at $q=1$; Figs. \ref{fig:other_hds} and \ref{fig:different_qs} investigate other $k$s and $q$s), with imaging and tabular datasets evaluated separately. 
As observed for MNIST, both LCR and VS were generally larger than the cluster count. 
The principal exceptions were the CIFAR image datasets, where the absence of meaningful substructure produced hundreds of very small clusters. 
Correlations between the number of clusters and LCR, and between the number of clusters and VS, were both low ($R^2=0.05$ and $0.06$, respectively), indicating that these entropy measures capture information not captured by clustering. 
In contrast, LCR and VS correlated strongly with each other for both imaging and tabular datasets ($R^2=1.00$ and $0.90$, respectively).
Fig. \ref{fig:different_qs} shows results for $q=0$ and $\infty$.

The picture changed when the similarity matrix $Z$ was scaled by varying the parameter $k$ in Eq. \ref{eq:ED}. 
Fig. \ref{fig:other_hds} shows results for tabular datasets for three $k$ values spanning 15 orders of magnitude---$2.6\times10^{-8}, 2.0\times10^{-1}$, and $8.4\times10^{7}$---selected to bracket the reciprocals of the minimal and maximal entries of the distance matrices. 
We refer to $k$ as the \textit{half-distance} (by analogy with half-life): consider elements $i$ and $j$ separated by a distance $d_{ij}$ with similarity $z_{ij}$; increasing their separation by an additional $k$ halves the similarity $z_{ij}/2$.

The half-distance $k$ has a major impact on the correlation between LCR and VS. When $k$ is very small, all $z_{ij}$ approach 1, collapsing the system to a single individual for both LCR and VS (Fig. \ref{fig:other_hds}a and c, left), yielding a perfect correlation ($R^2=1.00$).
When $k$ is very large, all off-diagonal $z_{ij}$ approach 0, making every element essentially unique; consequently, both LCR and VS converge to the number of elements in the dataset, and the correlation again is perfect ($R^2=1.00$) (Fig. \ref{fig:other_hds}a and c, right).
Only at intermediate $k$ values do LCR and VS provide independent information (Figs. \ref{fig:other_hds}a and \ref{fig:other_hds}c, middle).

To further illustrate this effect, Figs. \ref{fig:other_hds}b and \ref{fig:other_hds}d display $R^2$, the Pearson correlation, and the corresponding covariance  as functions of $k$ for imaging and tabular datasets.
For tabular datasets at $q=1$, the $R^2$ and correlation curves exhibit several relative minima; the absolute minima---$R^{2}=0.68$ and correlation $=0.83$---did not occur at the same $k$ value that yields the maximal covariance, near $k=0.2$. 
This $k$ value for the middle panels of Fig. \ref{fig:other_hds}a and c was chosen for this reason.
We observed different-shaped curves for $q=0$ and $q=\infty$, and the values of $k$ at the extrema had no obvious connection to each other or to the extreme regimes and also differed between the tabular and imaging datasets; it is unclear how to predict their location for the general case of a generic similarity matrix $Z$. 

\section{Discussion}
Quantifying the information contained in a system is essential for virtually any complex system.
This includes ML datasets, where the fact that most elements are generally unique---no repeated images (in image datasets) or rows (in tabular datasets)---renders traditional entropy (Shannon entropy and the other Rényi entropies) uninformative. 
This is because traditional entropy measures information encoded by the shape of the elements' frequency distribution, but when the elements each appear only once, this distribution is flat (Fig. \ref{fig:overview}): there is no additional information there.
Traditional entropy \textit{is} used to measure the distribution of class sizes, which often vary; this is class balance. However, there is no obvious way to use class balance to measure the information contained within or between classes, which is a property of the elements themselves.

\subsection{Sentropy as a measure of dataset composition}
One way to get at this missing information is to use clustering algorithms to quantify dataset substructure as the number of clusters and possibly clusters' relative sizes (which can be measured by traditional entropy).
Like LCR and VS, which require a choice of similarity measure and $q$, clustering algorithms require choices about parameters; for example, HDBSCAN requires choosing \texttt{min\_cluster\_size}, \texttt{min\_samples}, etc. 

Conceptually, the main difference between clustering and sentropy is that sentropy accounts for similarities both within and between clusters, whereas most clustering algorithms (and classifiers) lack this ability: effectively, the within-cluster (or intra-class) similarity is always 1 and the between-cluster similarity is always 0.
The exception is fuzzy clustering, in which each element can belong to more than one cluster, with each element being parameterized by its strength of membership for one or more clusters.
One can view LCR as carrying fuzzy-clustering to its logical limit, with each unique element being its own cluster and the similarities giving other elements' strength of membership in that cluster and thereby contributing to the size of the cluster.
This is a fuzzy-clustering interpretation of the ordinariness, $Z\mathbf{p}_i$: a big cluster to which many unique elements belong fairly strongly would be considered ``ordinary'' (Fig. \ref{fig:mnist}c). 
This perspective provides another way of seeing that LCR is the entropy of a system after accounting/adjusting/correcting for the similarity among its elements. 

Traditional entropy accounts for frequency; LCR accounts for both frequency and similarity. 
In this way, LCR is a richer, more complete measure of the information in a system.
Note that, regardless of whether they are represented as effective numbers (as we \cite{nguyen2026} and others \cite{hill1973diversity, jost2006entropy, leinsterEntropyDiversityAxiomatic2020} have advocated) or as bits of information, LCR is always bounded above by the corresponding traditional entropy \cite{leinsterEntropyDiversityAxiomatic2020}; for example, at $q=1$ LCR never exceeds Shannon entropy. 
Thus, similarity can be seen as sapping individual elements of some of their uniqueness.
However, in doing so, it reveals differences between datasets that traditional entropy cannot (Fig. \ref{fig:overview}). 

How do LCR and VS compare with the number of clusters (as identified by HDBSCAN)? 
Predicting the relationship is difficult: large clusters whose peripheral elements are relatively distinct tend to increase the LCR/VS values, whereas strong similarities among elements---both within a cluster and across neighboring clusters---tend to reduce them (see Figs.~\ref{fig:overview}, \ref{fig:trends}c). 
Indeed, many of the datasets we examined exhibited a pronounced difference in magnitude between tabular and imaging datasets when comparing the HDBSCAN cluster count with LCR and VS. 
For tabular data sets, both LCR and VS almost always surpassed the HDBSCAN cluster count, often by 1-2 orders of magnitude, reflecting that many rows remain unclustered by hard clustering (these do not contribute to the HDBSCAN total). 

In contrast, for imaging datasets, LCR and VS were typically 1-2 orders of magnitude \textit{lower} than the number of clusters, indicating substantial similarity among elements \textit{across} clusters, which hard clustering does not capture (Fig. \ref{fig:mnist}c). 
We interpret these systematic differences as evidence of the advantages of sentropy relative to hard clustering, for describing a system's structure. 
We do not regard this as a failure of HDBSCAN; a limitation of our study is that we did not explore a range of HDBSCAN hyperparameters, since the goal was simply to illustrate the general relationship between similarity-sensitive entropy and clustering. However, we note that by the ``eye test,'' UMAP plots validate HDBSCAN's clustering, suggesting hyperparameter search is unlikely to meaningfully affect the findings or conclusions presented.

\subsection{The effect of scaling $Z$}\label{sec:k}

We demonstrated that the relationship between LCR and VS depends strongly on how the similarity- matrix values are scaled, which we illustrated by introducing the half-distance parameter $k$.
Recall that similarity between two elements ranges from 0 to 1; scaling modifies this relationship by raising the values to a power, thereby shifting the curve in the middle of this range, while maintaining the anchoring at each end (at 0 and 1).

Why should one need to scale the similarity between elements?
Should scaling be permitted at all?
When similarity is derived from external knowledge or some principle or outside requirement, we argue that scaling should \textit{not} be applied: the external knowledge defines and thereby constrains the similarity function \textit{a priori}, its output for any given pair is fixed, and the similarity matrix simply collects these values for all pairs.
An example is binding similarity in immune repertoires, defined as the relative dissociation constants of two antibodies (or TCRs) for a given antigen (or set of antigens) \cite{aroraRepertoirescaleMeasuresAntigen2022}. This is the unique definition that preserves biophysical additivity; arbitrary scaling would break this relationship and thereby produce a similarity matrix that is objectively incorrect.

In contrast, when similarity is heuristic and not constrained by such knowledge, as with the Euclidean-\allowbreak distance-\allowbreak based similarity measures used here (across different embeddings), there is no \textit{a priori} reason to privilege the default similarity ($k=1$) over other values of $k$. For instance, RMSD provides a perfectly valid similarity definition and corresponds to Euclidean distance with $k=1/\sqrt{d}$, where $d$ is the dimension of the underlying data (number of columns for tabular data, number of pixels $\times$ number of color channels for image data, 2 for the umap embedding, etc).
Therefore, in heuristic cases, scaling can serve as a reasonable dial that investigators can adjust to obtain values that make the most sense given the context/are maximally informative about the system.
Such choices arise often in measurement. For example, the parameter $q$ is an analogous dial in traditional entropy (as well as sentropy), allowing investigators to adjust the emphasis on frequency.
As with $k$, prior information can constrain choice of $q$; for example, when the measurement has to do with collision frequency, $q=2$ should be used, consistent with $q=2$ corresponding to Simpson's index.
Whether specific values of $k$ might similarly be linked to known quantities of interest is a question left for future investigation.

A key feature of scaling is that it interpolates sentropy between 1 (for small $k$) and traditional entropy (for large $k$).
When $k$ is small, the off-diagonal entries of the similarity matrix approach 1, implying that all unique elements are mutually similar; in the limit the system behaves as if it contained a single unique element.
Conversely, for large $k$ the off-diagonal entries approach zero, yielding the identity matrix in the limit; the unique elements become completely dissimilar, and traditional entropy is recovered.
In a system where the $n$ elements are all unique---typically the case for ML data sets, including the imaging and tabular data sets examined here---the effective-number form of traditional entropy equals the dataset size.
Consequently, for such uniform systems $k$ interpolates both LCR and VS from 1 to $n$. Other scaling approaches may exhibit the same property.

Which $k$ should one choose? 
Qualitatively, the answer is a value roughly midway between the extremes (Fig. \ref{fig:other_hds}). 
However, our empirical studies indicate that the optimal choice is not obvious.
We found that the correlation between LCR and VS is minimized at $k=0.2$ for the tabular data sets and $3.4$ for the imaging data sets we tested---not at any obvious definition of the midpoint of the range (defined for example as the geometric mean of the $k$ values where the correlation first falls to 0.99 on each side).
Moreover, the shapes of these curves, with their multiple extrema, do not appear to follow a simple parametric form, making it difficult to predict precisely where the two metrics are maximally mutually informative.
We also observed that even at the $k$ that minimizes the LCR-VS correlation, the $R^2$ between LCR and VS remained relatively high in absolute terms, at $0.68$. Perhaps this was to be expected, since both metrics quantify the effective number of elements using the same underlying similarity measure, albeit in different ways (see Section \ref{sec:conclusion}). The dependence on $k$, however, was quite strong.

Our results suggest first testing whether the entropy lies near 1 or $n$; if so, to explore, if possible, different $k$ values to delineate a non-trivial range; and finally to select a $k$ approximately in the middle of this range (on a logarithmic scale).
Both the choice of similarity measure and the value of $k$ have effects on the resulting entropy that cannot be ignored.
Thus to some extent, from a quantitative perspective, the information in a system is in the eye of the beholder.

\section{Conclusion}\label{sec:conclusion}

So, which sentropy: LCR or VS? Our results support defaulting to LCR when the goal is to measure the similarity-adjusted entropy of a system, and to prefer VS only in one of two special cases: (\textit{i}) when the elements are usefully thought of as linear combinations of a set of mutually orthogonal ur-elements/eigen-elements, or (\textit{ii}) when the system possesses a quantum-mechanical character (see below).
If the aim is a more general multidimensional characterization, there is no clear reason to avoid employing both metrics; keeping in mind that they are likely to correlate, our findings recommend selecting $k$ within the non-trivial range (see Section \ref{sec:k}) to maximize their independence and thereby their combined informativeness.

We note that, quantitatively, VS $\geq$ LCR for any given $k$ and $q$.
We proved this relationship for all $q\geq0$ and for $q<0$ when $Z$ is full rank.
This relationship can be rationalized by recognizing that, although VS is a similarity-sensitive entropy, its computation includes a traditional entropy component: after incorporating similarity to derive the eigen-elements, VS is the traditional entropy of the resulting eigenvalues; traditional entropy is always greater than or equal to similarity-sensitive entropy.

\begin{figure}
    \centering
    \includegraphics[width=0.7\linewidth]{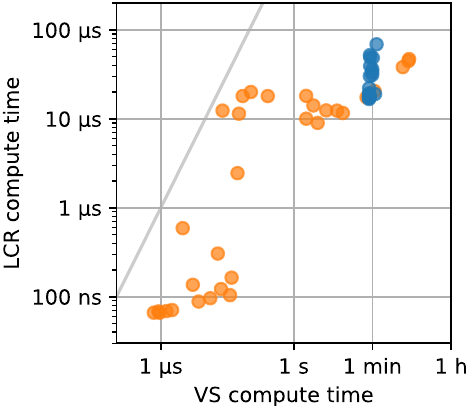}
    \caption{Compute times for all 53 datasets for LCR (using the \texttt{sentropy} Python package \cite{nguyen2026}) vs. VS (using \texttt{vendi-score}). Imaging datasets (blue) were all 10,000 images. Smaller datasets were faster. Note the different timescales.}
    \label{fig:times}
\end{figure}

From a practical perspective, LCR has three advantages. First, its time complexity is $\mathcal{O}{(n^2)}$ whereas computing eigenvalues---which is required to obtain the VS---has time complexity $\mathcal{O}{(n^3)}$ \cite{friedman2022vendi}. Empirically, we observed a speed advantage of the \texttt{sentropy} package over the \texttt{vendi-score} package even when all elements were unique, as is often the case in ML datasets (Fig. \ref{fig:times}). In addition, computing the LCR is less computationally expensive whenever elements appear multiple times (i.e., when $\mathbf{p}$ is not flat/uniform), because the similarity matrix for LCR has order equal to the number of \textit{unique} elements, whereas VS requires a matrix whose order equals the \textit{total} number of elements. 

Second, LCR does not require the similarity matrix to be PSD, whereas VS does.
Similarity matrices need not be PSD; a perfectly reasonable set of similarity measurements can produce a matrix that fails the PSD condition, which would complicate the use of VS.
Even symmetric similarity matrices---the most common and most natural type, satisfying $z_{ij}=z_{ji}$, so that similarity is reciprocal---are not guaranteed to be PSD, even when their entries obey the triangle inequality.
The prevalence of such non-PSD cases in practice is unknown; we note that all similarity matrices examined for our datasets were PSD (a consequence of how we chose our similarity measures; this follows from Theorem 2.11 in \cite{leinster2013magnitude}).
Thus, PSD-ness is irrelevant for LCR but must be checked to apply VS.

And third, LCR in its various forms \cite{chao2010phylogenetic, chaoAttributediversityApproachFunctional2019} (Table \ref{table:entropies}) is part of an extensively characterized framework \cite{leinsterEntropyDiversityAxiomatic2020} that provides similarity-sensitive analogs of mutual information, relative entropy, and cross-entropy and offers methods for partitioning the overall similarity-sensitive entropy ($\gamma$) into within-class ($\alpha$) and between-class ($\beta$) components.  
In entropic terms, $\alpha$ is the (effective-number form of the) joint entropy of the elements together with their class label; $\beta$ is the mutual information between the elements and the class label; and $\gamma$ is the marginal entropy of the elements irrespective of class.  
In this work we computed LCR in its $\gamma$ form; when only a single class is present, $\gamma$ coincides with $\alpha$.  Measuring the full $\alpha$, $\beta$, and $\gamma$ decomposition across ML classes is a promising direction for future research.  (The \texttt{sentropy} package computes $\alpha$, $\beta$, and $\gamma$.)

The necessary mathematics for extending this decomposition to VS exists, because VS is based on von Neumann (quantum) entropy, for which quantum mutual information, quantum relative entropy, and related quantum quantities are well understood \cite{nielsen2010quantum}.
However, computing these quantities requires a matrix that simultaneously encodes each element and its class membership.
If there are $n$ total elements and $k$ classes, the VS decomposition demands either an $nk \times nk$ matrix or an $n \times k \times n \times k$ 4-dimensional tensor (which can be reduced to $n \times n$ and $k \times k$ matrices by taking a partial trace).  
In contrast, LCR requires only the similarity matrix $Z$.  
Consequently, VS needs an explicit notion of similarity both between classes and between elements, whereas in LCR any class-level similarity arises purely from element-level similarities within the classes.  
In principle, the VS formulation would permit the confusing situation in which two classes contain exactly the same individuals but nonetheless have zero similarity---a scenario that cannot occur under LCR, in which the similarity between classes is purely a function of similarities among the elements they contain.

In sum, LCR and VS provide complementary descriptions of similarity-sensitive entropy, a unifying concept that is likely to find applications in many domains in which traditional entropy is currently a mainstay. We anticipate that the deeper understanding of their relationship and properties will aid such investigations.

\section{Acknowledgements}
This work was supported by the Gordon and Betty Moore Foundation and by the NIH under grants R01HL150394, R01HL150394-SI, R01AI148747, and R01AI148747-SI.

\appendix
\section{Comparison to 3D UMAP embedding}\label{app:randomseeds}

\begin{figure}[h!]
    \centering
    \includegraphics[width=0.5\linewidth]{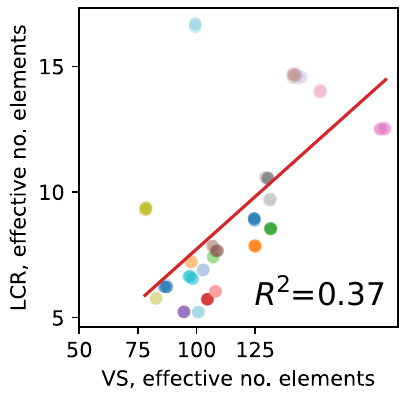}
    \caption{Results from 10 replicates with different random seeds for each dataset. Legend as in Fig. \ref{fig:trends}a.}
    \label{fig:3D_umap}
\end{figure}

Fig. \ref{fig:3D_umap} shows results of the same analysis as in the right-hand panel of Fig. \ref{fig:trends}a but with the UMAP embeddings in three dimensions instead of two. This supports the lower-dimensional observations that VS and LCR tend to correlate to each other, but do contain independent information ($R^2=0.37$). 

\section{Other mathematical results}

\Cref{lem: R >= Z} has an interesting consequence which is not used in our proof of \Cref{thm: Vendi bounds LCR}, namely
\begin{cor}\label{cor: row sums >= evals}
    Let Z, $\lambda_i$, and $r_i$ be as in \Cref{lem: R >= Z}. Denote by $r_{(i)}$ the $i$th row-sum of $Z$ ordered from largest to smallest, so that e.g. $r_{(1)} = \max_i r_i$. Then
    $r_{(i)} \geq \lambda_i$ for all $i$ in $\{1, \ldots, n\}$
\end{cor}
\begin{proof}
    This is an immediate consequence of \Cref{lem: R >= Z} and the Weyl monotonicity theorem \cite{Weyl:1912}. 
\end{proof}
This would seem a good basis for proving \Cref{thm: Vendi bounds LCR} except $VS_q$ depends on $\sum_i \lambda_i^q$, but $D_q\left(Z, \frac{1}{n}\right)$ depends on $\sum_i r_i^{q-1}$ rather than $\sum_i r_i^{q}$. It is clear, however, that \Cref{cor: row sums >= evals} leads to some bound involving $VS_q$ and $D_{q+1}$. Further computation reveals that the relevant bound is
\begin{cor}\label{cor: powers of evals and row sums}
    \begin{align}
        \left[VS_q(Z) \right]^{1-q} \leq n^{\frac{1+q}{q}} [D_{q+1}]^{-q}
    \end{align}
    for $q>0$;
    \begin{align}
        VS_q(Z) \begin{cases}
            \leq n^{\frac{1+q}{q(1-q)}} [D_{q+1}]^{-\frac{q}{1-q}} \text{ for } 0<q<1\\
            \geq n^{\frac{1+q}{q(1-q)}} [D_{q+1}]^{-\frac{q}{1-q}} \text{ for } q>1
        \end{cases}
    \end{align}
\end{cor}
\begin{proof}
    This is a direct consequence of \Cref{cor: row sums >= evals}. Since $r_{(i)} \geq \lambda_i$, and since $x\rightarrow x^q$ is monotonically increasing for $q>0$, we have for each $i$ that
    \begin{gather}
        r_{(i)}^q \geq \lambda_i^q\\
        \Rightarrow\\
        \sum_i \lambda_i^q \leq \sum_i r_i^q
    \end{gather}
    However, for $q\neq 1$, 
    \begin{align}
        \sum_i r_i^q &= n^{1+q} [D_{q+1}]^{-q} \\
        \sum_i \lambda_i^q &= n^q \left[VS_q(Z) \right]^{1-q}
    \end{align}
    thereby establishing the claim for $q\neq 1$. The $q=1$ case then follows by continuity in $q$.
\end{proof}
Note that this bound is weaker than \Cref{lem: lem 1} for $q>1$, since $r_i^{q-1} \leq r_i^q$. On the other hand, for $0<q<1$ it provides an upper (rather than lower) bound on $VS_q(Z)$, since $1-q>0$, but this bound is also weaker than this simple bound that $VS_q(Z)<n$. As an illustrative example, consider the case of $q=\frac{1}{2}$. Then \Cref{cor: powers of evals and row sums} leads to
\begin{align}
    VS_{\frac{1}{2}} \leq \frac{n^6}{D_{\frac{3}{2}}\left(Z, \frac{1}{n}\right)}.
\end{align}
This is clearly much weaker than $VS_{\frac{1}{2}}<n$, since $D_{\frac{3}{2}}\left(Z, \frac{1}{n}\right)\leq n$ means the right hand side is never smaller than $n^5$. In fact, this is in a sense the best case. Taking the inequality at general $0<q<1$, namely
\begin{align}
    VS_q(Z) \leq n^{\frac{1+q}{q(1-q)}} [D_{q+1}]^{-\frac{q}{1-q}},
\end{align}
$D_{q+1} \leq n$ means the right hand side is bounded below by $n^{\frac{1 + q - q^2}{q (1-q)}}$. Minimizing the exponent (with the restriction that $0<q<1$) gives $q = \frac{1}{2}$, so this right hand side is always bounded below by $n^5$, where as the left hand side is bounded above by $n$---a very weak bound.

Another bound can be proven:
\begin{lem}\label{lem: inequality 2}
    Let $Z$, $\text{VS}_{q}(Z)$, and $D_q(Z, \frac{1}{n})$ be as in \Cref{thm: Vendi bounds LCR}. Then 
    \begin{equation}
        \mathrm{VS}_{1}(Z) \leq \frac{1}{n} \left( \sum_{i,j} Z_{ij} \right) D_{1}(Z, \frac{1}{n})
    \end{equation}
\end{lem}
\begin{proof}
    We begin by constructing the diagonal matrix $W$ according to 
    \begin{align}
        W_{ii} &= \frac{1}{T}\sum_j Z_{ij}\\
        T &= \sum_{i,j} Z_{ij}
    \end{align}
    Notice that $W$ is positive definite (since the entries of $Z$ are non-negative and the diagonal entries are $1$) and $\tr W = 1$. Treating $W$ and $Z/n$ as quantum density matrices, the quantum relative entropy between them should be positive, i.e.
    \begin{gather}
        \tr \left[ \frac{Z}{n} \left( \log \frac{Z}{n} - \log W \right)  \right] \geq 0 \\
        \Rightarrow \nonumber\\
        \log \mathrm{VS}_{q=1} (Z) = - \tr \frac{Z}{n} \log \frac{Z}{n} \leq - \tr \frac{Z}{n} \log W.
    \end{gather}
    However, since $W$ is diagonal we have that
    \begin{align}
        - \tr \frac{Z}{n} \log W &= - \sum_{i} \frac{Z_{ii}}{n} \log W_{ii} \\
        &= - \sum_{i} \frac{1}{n} \log \frac{\sum_j Z_{ij}}{T} \\
        &= - \sum_{i} \frac{1}{n} \left( \log \frac{\sum_j Z_{ij}}{n} -\log \frac{T}{n}\right)\\
        &= \left(- \sum_{i} \frac{1}{n} \log \frac{\sum_j Z_{ij}}{n}\right) + \log \frac{T}{n}
    \end{align}
    As such, we have that
    \begin{align}
        \log \mathrm{VS}_{1} (Z) \leq  \left(- \sum_{i} \frac{1}{n} \log \frac{\sum_j Z_{ij}}{n}\right) + \log \frac{T}{n}
    \end{align}
    Exponentiating both sides, and recognizing that $e^{- \sum_{i} \frac{1}{n}  \log \frac{\sum_j Z_{ij}}{n}} = D_{1}(Z, \frac{1}{n})$, we have that
    \begin{align}
        \mathrm{VS}_{1} (Z) \leq \frac{T}{n} D_{1}(Z, \frac{1}{n})
    \end{align}
    proving the theorem.

\end{proof}

\section{Notes on previous versions and alternative proofs}

\subsection{Proof that $VS\geq LCR$ for $q \in \{2,3\}$  }\label{a:proofqs}

In a previous version of this paper, \Cref{thm: Vendi bounds LCR} was left as a conjecture except for $q\in{2,3,+\infty}$, and $q\in[-\infty, 0]$ when $Z$ is full rank. The proofs for $q=+\infty$ and for $q\in[-\infty, 0]$  were as currently presented in \Cref{sec: math}. The previous proofs for $q \in \{2,3\}$ are presented below
\begin{proof}[Proof of \Cref{thm: Vendi bounds LCR} in the special cases of $q=2$ and $q=3$]
    We will consider the cases separately. For convenience, in what follows, denote by $\lambda_i$ the eigenvalues of $Z$ and $r_i$ the row sums of $Z$, i.e. $r_i = \sum_j Z_{ij}$. 
    
    At $q=2$,
    \begin{align}
        \mathrm{VS}_2(Z) &= e^{- \log \tr \left(\frac{Z}{n}\right)^2} = \frac{1}{\tr \left(\frac{Z}{n}\right)^2} = \frac{n^2}{\sum_{i,j} Z^2_{i,j}}
    \end{align}
    However, since $0 \leq Z_{ij} \leq 1 \forall (i, j)$, it follows that $Z_{i,j}^2 \leq Z_{ij}$ and so $\displaystyle\sum_{i,j} Z_{i,j}^2 \leq \displaystyle\sum_{i,j} Z_{ij}$.
    Therefore,
    \begin{align}
        \frac{n^2}{\sum_{i,j} Z^2_{i,j}} \geq \frac{n^2}{\sum_{i,j} Z_{i,j}}
    \end{align}
    However, $\frac{n^2}{\sum_{i,j} Z_{i,j}}$ is simply $D_{2}(Z, \frac{1}{n})$, proving the $q=2$ case.

    For $q=3$, we have that
    \begin{align}
        \tr \left(Z^3\right) = \sum_{i,j} \left( Z^2\right)_{ij} Z_{ij} \leq \sum_{i,j} \left( Z^2\right)_{ij}\\
        = \sum_{i,j,k} Z_{ik} Z_{kj} = \sum_k \left( \sum_i Z_{ki} \right) \left( \sum_j Z_{kj} \right) = \sum_i r_i^2.
    \end{align}
    This gives
    \begin{align}
        \mathrm{VS}_3(Z) = e^{- \frac{1}{2} \log \tr \left(\frac{Z}{n}\right)^3} = \sqrt{\frac{n^3}{\tr \left(Z^3\right)}} \geq \sqrt{\frac{n^3}{\sum_i r_i^2}} = D_{3}(Z, \frac{1}{n})
    \end{align}
    proving the $q=3$ case.
\end{proof}

\subsection{Hoffman 1966 and a proof for integer $q\geq 2$}\label{a:approach}

A previous version of this paper conjectured the following result, noting that it would help in proving \Cref{lem: lem 1} (which was also stated as a conjecture).
\begin{thm}[Hoffman 1966]\label{thm: matrix ineq}
    \label{thm: sum of rows of power vs power of sum of rows}
    Denote by $\mathbf{1}$ the $n$ dimensional vector each of whose entries are $1$, Let $M$ be any $n \times n$ symmetric matrix with non-negative entries, and denote by $r = M \mathbf{1}$ (i.e. $r_i=\sum_j M_{ij}$) the row-sums of $M$. Then for any integer $\alpha \geq 0$
    \begin{align}
        \mathbf{1}^\intercal M^\alpha \mathbf{1}  \leq \mathbf{1}^\intercal \left(M \mathbf{1}\right)^\alpha
    \end{align} 
\end{thm}
Since then we have learned that this result was proved by Hoffman in \cite{hoffman1966} following up on earlier work by London in \cite{london1966}, which had proved the special cases of $n\leq 3$ for all integers $\alpha \geq 1$, and $5 \geq \alpha \geq 1$ for arbitrary $n>0$. This result leads to an alternative proof of \Cref{lem: lem 1} for integer $q\geq 1$ (and thus a proof of \Cref{thm: Vendi bounds LCR} for integer $q\geq 2$), which is essentially what we had in mind in the earlier version of this work. This proof is as follows:
\begin{proof}[Proof of \Cref{lem: lem 1} in the special case of integer $q\geq 1$]
    Consider that
    \begin{align}
        \tr \left( Z^{q} \right) = \sum_{i,j} Z_{ij} \left(Z^{q-1} \right)_{i,j} \leq \sum_{i,j} \left(Z^{q-1} \right)_{i,j} = \mathbf{1}^\intercal Z^{q-1} \mathbf{1}
    \end{align}
    since $Z$ is symmetric, $0\leq Z_{i,j}\leq 1$ for each $i,j$, and $Z^{q-1}$ is non-negative (since it is the product of $q-1$ non-negative matrices). But since $Z$ is symmetric and non-negative, and by assumption $q-1\geq 0$ is an integer, by \Cref{thm: matrix ineq}
    \begin{equation}
        \mathbf{1}^\intercal Z^{q-1} \mathbf{1} \leq \mathbf{1}^\intercal \left(Z \mathbf{1}\right)^{q-1}
    \end{equation}
\end{proof}

\subsection{Python code to search for counterexamples}\label{a:code}

In a previous version of this paper, we discussed a numerical search for counterexamples to \Cref{thm: Vendi bounds LCR} (partly inspred by \cite{Couch:2023pav}). The code used for that search is provided below.

Note this does not search over the whole space of non-negative PSD similarity matrices. Rather, by construction, it only searches over completely positive matrices. For $n$ by $n$ matrices with $n\leq 4$, all non-negative PSD matrices are completely positive, but for $n\geq 5$ this is not the case \cite{maxfield1962}. Given the improvements in what we have proven over that version, we no longer see a need for such a numerical counterexample search; the code is provided only for archival reasons. 

\begin{lstlisting}[language=Python, basicstyle=\small, frame=double]
import numpy as np
import pandas as pd
from scipy.optimize import minimize
from tqdm import tqdm, trange
from multiprocessing import Pool

n_workers = 96

def uniform_div(Z,q): 
    p = (1.0/Z.shape[0])*np.ones_like(Z[0])
    Zp = Z@p
    if q==1:
        h1 = -np.sum(p*np.log(Zp))
        return np.exp(h1)
    else:
        X = np.sum(p*(Zp**(q-1)))
        h = (1/(1-q)) * np.log(X)
        return np.exp(h)

def vendi_score(Z,q):
    evals = np.linalg.eigvalsh(Z/Z.shape[0])
    # keep only the support
    evals = np.array(
        [e for e in evals if e>0]
    ) 
    if q==1:
        h1 = -np.sum(evals*np.log(evals))
        return np.exp(h1)
    else:
        X = np.sum(evals**q)
        h = (1/(1-q)) * np.log(X)
        return np.exp(h)

def func(x, n, r, q):
    Z = x_to_Z(x, n, r)
    # Compute vendi score
    ret = vendi_score(Z, q)
    # Compute and subtract 
    # uniform diversity
    ret -= uniform_div(Z, q)
    return ret

def x_to_Z(x, n, r):
    # map from reals to positive
    # reals, reshape
    w = (x**2).reshape(r, n)
    # normalize rows
    w = w/np.sqrt((w**2).sum(0))
    Z = w.transpose()@w 
    return Z

def find_min(n, r, q, scale):
    args = (n, r, q)
    x0 = scale*np.random.normal(0, 1, n*r)
    min_obj = minimize(func, x0, args)
    ret = dict(min_obj)
    ret.update({
        'x0': x0, 
        'Z0': x_to_Z(x0, n, r), 
        'Z': x_to_Z(min_obj.x, n, r),
    })
    return ret

qlist = [
    -10, 
    -2.7,
    -2, 
    -1, 
    -0.3, 
    0, 
    0.3, 
    1, 
    2.7, 
    3.4, 
    4, 
    5, 
    2*np.pi, 
    7, 
    8, 
    9.1, 
    10,
]
sizes = [3, 5, 10, 20]
scales = [0.1, 1, 2, 10]
replicates = [i for i in range(1,6)]
combos = [{
    'q': q, 
    'n': n, 
    'r': r, 
    'rep': rep, 
    'scale': s,} 
for q in qlist for n in sizes for s in scales 
for rep in replicates 
for r in set([x for x in [2, n//2, n] if x>1])]


def map_func(combo):
    q = combo['q']
    r = combo['r']
    n = combo['n']
    rep = combo['rep']
    scale = combo['scale']
    row = combo
    try:
        row.update(find_min(n, r, q, scale))
        return row
    except np.linalg.LinAlgError:
        return row   

results = []
np.random.shuffle(combos)

n_batches = len(combos)//n_workers
if n_batches*n_workers< len(combos):
    n_batches +=1


pool = Pool(n_workers)
outfile = 'counterexample_search.pkl'

for batch in trange(n_batches):
    ix_start = batch*n_workers
    ix_end = ix_start + n_workers
    rows = pool.map(
        map_func, 
        combos[ix_start:ix_end],
    )
    results = results + rows 
    results_df = pd.DataFrame(results)
    results_df.to_pickle(outfile)
\end{lstlisting}

\bibliographystyle{JHEP_edited}
\bibliography{bibliography}

@misc{knuth2026claude,
  author      = {Knuth, Donald E.},
  title       = {Claude's Cycles},
  year        = {2026},
  month       = feb,
  note        = {Revised 06 March 2026},
  url         = {https://www-cs-faculty.stanford.edu/~knuth/papers/claude-cycles.pdf},
  howpublished = {\url{https://www-cs-faculty.stanford.edu/~knuth/papers/claude-cycles.pdf}},
  institution = {Stanford University, Department of Computer Science}
}

@article{chinnENRICHingMedicalImaging2023,
	title = {{ENRICHing} medical imaging training sets enables more efficient machine learning},
	volume = {30},
	issn = {1527-974X},
	doi = {10.1093/jamia/ocad055},
	language = {eng},
	number = {6},
	journal = {Journal of the American Medical Informatics Association: JAMIA},
	author = {Chinn, Erin and Arora, Rohit and Arnaout, Ramy and Arnaout, Rima},
	month = may,
	year = {2023},
	pmid = {37036945},
	pmcid = {PMC10198519},
	pages = {1079--1090},
}

@article{visipediaInat_comp2017READMEmd,
	title = {inat\_comp/2017/{README}.md at master · visipedia/inat\_comp},
	url = {https://github.com/visipedia/inat_comp/blob/master/2017/README.md},
	language = {en},
	urldate = {2025-08-10},
	journal = {GitHub},
	author = {{visipedia}},
}

@inproceedings{wangChestXRay8HospitalScaleChest2017,
	title = {{ChestX}-{Ray8}: {Hospital}-{Scale} {Chest} {X}-{Ray} {Database} and {Benchmarks} on {Weakly}-{Supervised} {Classification} and {Localization} of {Common} {Thorax} {Diseases}},
	shorttitle = {{ChestX}-{Ray8}},
	doi = {10.1109/CVPR.2017.369},
	booktitle = {2017 {IEEE} {Conference} on {Computer} {Vision} and {Pattern} {Recognition} ({CVPR})},
	author = {Wang, X. and Peng, Y. and Lu, L. and Lu, Z. and Bagheri, M. and Summers, R. M.},
	month = jul,
	year = {2017},
	note = {ISSN: 1063-6919},
	pages = {3462--3471},
    eprint={1705.02315},
    archivePrefix={arXiv},
    primaryClass={cs.CV},
    url={https://arxiv.org/abs/1705.02315}, 
}

@article{mcinnesHdbscanHierarchicalDensity2017,
	title = {hdbscan: {Hierarchical} density based clustering},
	volume = {2},
	issn = {2475-9066},
	shorttitle = {hdbscan},
	url = {https://joss.theoj.org/papers/10.21105/joss.00205},
	doi = {10.21105/joss.00205},
	language = {en},
	number = {11},
	urldate = {2025-10-26},
	journal = {Journal of Open Source Software},
	author = {McInnes, Leland and Healy, John and Astels, Steve},
	month = mar,
	year = {2017},
	pages = {205},
}

@incollection{renyiMeasuresEntropyInformation1961,
	title = {On {Measures} of {Entropy} and {Information}},
	volume = {4.1},
	url = {https://projecteuclid.org/ebooks/berkeley-symposium-on-mathematical-statistics-and-probability/Proceedings-of-the-Fourth-Berkeley-Symposium-on-Mathematical-Statistics-and/chapter/On-Measures-of-Entropy-and-Information/bsmsp/1200512181},
	urldate = {2025-09-29},
	booktitle = {Proceedings of the {Fourth} {Berkeley} {Symposium} on {Mathematical} {Statistics} and {Probability}, {Volume} 1: {Contributions} to the {Theory} of {Statistics}},
	publisher = {University of California Press},
	author = {Rényi, Alfréd},
	month = jan,
	year = {1961},
	pages = {547--562},
}

@article{chaoAttributediversityApproachFunctional2019,
	title = {An attribute-diversity approach to functional diversity, functional beta diversity, and related (dis)similarity measures},
	doi = {10.1002/ecm.1343},
	journal = {Ecological Monographs},
	author = {Chao, Anne and Chiu, Chun-Huo and Villéger, Sébastien and Sun, I Fang and Thorn, Simon and Lin, Yiching and Chiang, Jyh-Min and B. Sherwin, William},
	year = {2019},
}

@article{Shannon1951, 
  title={Prediction and Entropy of Printed English}, 
  author={Shannon, Claude E.}, 
  journal={Bell System Technical Journal}, 
  year={1951}
}

@article{shannon1948mathematical,
  title={A mathematical theory of communication},
  author={Shannon, Claude E},
  journal={The Bell system technical journal},
  volume={27},
  number={3},
  pages={379--423},
  year={1948},
  publisher={Nokia Bell Labs}
}

@article{tawilCompetitionCaliforniasMediCal2022,
	title = {Competition in {California}'s {Medi}-{Cal} {Managed} {Care} {Market} {Assessed} by {Herfindahl}-{Hirschman} {Index}},
	volume = {59},
	issn = {1945-7243},
	doi = {10.1177/00469580221127063},
	language = {eng},
	journal = {Inquiry: A Journal of Medical Care Organization, Provision and Financing},
	author = {Tawil, Michael and DiGiorgio, Anthony M.},
	year = {2022},
	pmid = {36168304},
	pmcid = {PMC9520176},
	pages = {469580221127063},
}

@article{aroraRepertoirescaleMeasuresAntigen2022,
	title = {Repertoire-scale measures of antigen binding},
	volume = {119},
	issn = {0027-8424, 1091-6490},
	url = {https://pnas.org/doi/full/10.1073/pnas.2203505119},
	doi = {10.1073/pnas.2203505119},
	language = {en},
	number = {34},
	urldate = {2022-08-20},
	journal = {Proceedings of the National Academy of Sciences},
	author = {Arora, Rohit and Arnaout, Ramy},
	month = aug,
	year = {2022},
	pages = {e2203505119},
}

@article{arnaoutFutureBloodTesting2021,
	title = {The {Future} of {Blood} {Testing} {Is} the {Immunome}},
	volume = {12},
	issn = {1664-3224},
	url = {https://www.frontiersin.org/articles/10.3389/fimmu.2021.626793/full},
	doi = {10.3389/fimmu.2021.626793},
	language = {en},
	urldate = {2022-08-20},
	journal = {Frontiers in Immunology},
	author = {Arnaout, Ramy A. and Prak, Eline T. Luning and Schwab, Nicholas and Rubelt, Florian and {the Adaptive Immune Receptor Repertoire Community}},
	month = mar,
	year = {2021},
	pages = {626793},
}

@book{leinsterEntropyDiversityAxiomatic2020, 
    place={Cambridge}, 
    title={Entropy and Diversity: The Axiomatic Approach}, 
    publisher={Cambridge University Press}, 
    author={Leinster, Tom}, 
    year={2021},
    eprint={2012.02113},
    archivePrefix={arXiv},
    primaryClass={q-bio.PE},
    url={https://arxiv.org/abs/2012.02113}, 
}

@article{Montemurro1994, 
  title={Entropy of natural languages: Theory and experiment}, 
  author={Montemurro, M. A. and Zanette, D. H.}, 
  journal={Scientific Programming}, 
  year={1994}
}

@article{hill1973diversity,
  title={Diversity and evenness: a unifying notation and its consequences},
  author={Hill, Mark O},
  journal={Ecology},
  volume={54},
  number={2},
  pages={427--432},
  year={1973},
  publisher={Wiley Online Library}
}

@article{jost2006entropy,
  title={Entropy and diversity},
  author={Jost, Lou},
  journal={Oikos},
  volume={113},
  number={2},
  pages={363--375},
  year={2006},
  publisher={Wiley Online Library}
}

@article{leinster2012measuring,
  title={Measuring diversity: the importance of species similarity},
  author={Leinster, Tom and Cobbold, Christina A},
  journal={Ecology},
  volume={93},
  number={3},
  pages={477--489},
  year={2012},
  publisher={Wiley Online Library}
}

@article{reeve2014partition,
      title={How to partition diversity}, 
      author={Richard Reeve and Tom Leinster and Christina A. Cobbold and Jill Thompson and Neil Brummitt and Sonia N. Mitchell and Louise Matthews},
      year={2016},
      eprint={1404.6520},
      archivePrefix={arXiv},
      primaryClass={q-bio.QM},
      url={https://arxiv.org/abs/1404.6520}, 
}

@article{chao2010phylogenetic,
  title={Phylogenetic diversity measures based on Hill numbers},
  author={Chao, Anne and Chiu, Chun-Huo and Jost, Lou},
  journal={Philosophical Transactions of the Royal Society B: Biological Sciences},
  volume={365},
  number={1558},
  pages={3599--3609},
  year={2010},
  publisher={The Royal Society}
}

@article{tsallis1988possible,
  title={Possible generalization of Boltzmann-Gibbs statistics},
  author={Tsallis, Constantino},
  journal={Journal of statistical physics},
  volume={52},
  number={1},
  pages={479--487},
  year={1988},
  publisher={Springer}
}

@article{couch2024beyond,
  title={Beyond Size and Class Balance: Alpha as a New Dataset Quality Metric for Deep Learning}, 
  author={Josiah Couch and Rima Arnaout and Ramy Arnaout},
  year={2024},
  eprint={2407.15724},
  archivePrefix={arXiv},
  primaryClass={cs.CV},
  url={https://arxiv.org/abs/2407.15724}, 
}

@article{couch2025x,
      title={X-Factor: Quality Is a Dataset-Intrinsic Property}, 
      author={Josiah Couch and Miao Li and Rima Arnaout and Ramy Arnaout},
      year={2025},
      eprint={2505.22813},
      archivePrefix={arXiv},
      primaryClass={cs.LG},
}

@article{friedman2022vendi,
  title={The Vendi Score: A Diversity Evaluation Metric for Machine Learning},
  author={Friedman, Dan and Dieng, Adji Bousso},
  journal={Transactions on Machine Learning Research},
  issn={2835-8856},
  year={2023},
  eprint={2210.02410},
  archivePrefix={arXiv},
  primaryClass={cs.LG},
  url={https://arxiv.org/abs/2210.02410}, 
}

@article{pasarkar2025vendiscope,
      title={The Vendiscope: An Algorithmic Microscope For Data Collections}, 
      author={Amey P. Pasarkar and Adji Bousso Dieng},
      year={2025},
      eprint={2502.10828},
      archivePrefix={arXiv},
      primaryClass={cs.LG},
      url={https://arxiv.org/abs/2502.10828}, 
}

@inproceedings{pasarkar2023cousins,
  title={Cousins Of The Vendi Score: A Family Of Similarity-Based Diversity Metrics For Science And Machine Learning},
  author={Pasarkar, Amey P and Dieng, Adji Bousso},
  booktitle={International Conference on Artificial Intelligence and Statistics},
  pages={3808--3816},
  year={2024},
  eprint={2310.12952},
  archivePrefix={arXiv},
  primaryClass={cs.LG},
  organization={PMLR}
}

@article{leinster2013magnitude,
  title={The magnitude of metric spaces},
  author={Leinster, Tom},
  journal={Documenta Mathematica},
  volume={18},
  pages={857--905},
  year={2013},
  eprint={1012.5857},
  archivePrefix={arXiv},
  primaryClass={math.MG},
  url={https://arxiv.org/abs/1012.5857}, 
}

@article{leinster2016maximizing,
  title={Maximizing diversity in biology and beyond},
  author={Leinster, Tom and Meckes, Mark W},
  journal={Entropy},
  volume={18},
  number={3},
  pages={88},
  year={2016},
  publisher={MDPI},
  eprint={1512.06314},
  archivePrefix={arXiv},
  primaryClass={cs.IT},
  url={https://arxiv.org/abs/1512.06314}, 
}

@article{vendi-score,
  author       = {Dan Friedman},
  title        = {vendi-score: A Diversity Metric for Machine Learning (Python package)},
  howpublished = {\url{https://pypi.org/project/vendi-score/}},
  version      = {0.0.3},
  year         = {2022},
  note         = {Accessed: 2025-10-13}
}

@article{Couch:2023pav,
    author = "Couch, Josiah and Nguyen, Phuc and Racz, Sarah and Stratis, Georgios and Zhang, Yuxuan",
    title = "{Possibility of entanglement of purification to be less than half of the reflected entropy}",
    eprint = "2309.02506",
    archivePrefix = "arXiv",
    primaryClass = "quant-ph",
    doi = "10.1103/PhysRevA.109.022426",
    journal = "Phys. Rev. A",
    volume = "109",
    number = "2",
    pages = "022426",
    year = "2024"
}

@book{nielsen2010quantum,
  title={Quantum Computation and Quantum Information: 10th Anniversary Edition},
  author={Nielsen, M.A. and Chuang, I.L.},
  isbn={9781139495486},
  url={https://books.google.com/books?id=-s4DEy7o-a0C},
  year={2010},
  publisher={Cambridge University Press}
}

@article{nguyen2026,
      title={$\textit{sentropy}$: A Python Package for Revealing Hidden Differences in Complex Datasets}, 
      author={Phuc Nguyen and Rohit Arora and Elliot D. Hill and Jasper Braun and Alexandra Morgan and Liza M. Quintana and Gabrielle Mazzoni and Ghee Rye Lee and Rima Arnaout and Ramy Arnaout},
      year={2026},
      eprint={2401.00102},
      archivePrefix={arXiv},
      primaryClass={q-bio.QM},
      url={https://arxiv.org/abs/2401.00102}, 
}

@article{lucieImprovedPython2024,
      title={$\textit{lucie}$: An Improved Python Package for Loading Datasets from the UCI Machine Learning Repository}, 
      author={Kenneth Ge and Phuc Nguyen and Ramy Arnaout},
      year={2024},
      eprint={2410.09119},
      archivePrefix={arXiv},
      primaryClass={cs.LG},
      url={https://arxiv.org/abs/2410.09119}, 
}

@article{mcinnesUMAPUniformManifold2020,
      title={UMAP: Uniform Manifold Approximation and Projection for Dimension Reduction}, 
      author={Leland McInnes and John Healy and James Melville},
      year={2020},
      eprint={1802.03426},
      archivePrefix={arXiv},
      primaryClass={stat.ML},
      url={https://arxiv.org/abs/1802.03426}, 
}

@inproceedings{torchvision,
    author = {Marcel, S\'{e}bastien and Rodriguez, Yann},
    title = {Torchvision the machine-vision package of torch},
    year = {2010},
    isbn = {9781605589336},
    publisher = {Association for Computing Machinery},
    address = {New York, NY, USA},
    url = {https://doi.org/10.1145/1873951.1874254},
    doi = {10.1145/1873951.1874254},
    booktitle = {Proceedings of the 18th ACM International Conference on Multimedia},
    pages = {1485–1488},
    numpages = {4},
    location = {Firenze, Italy},
    series = {MM '10}
}

@misc{mnist,
  title={MNIST handwritten digit database, 1998},
  author={LeCun, Yann and Cortes, Corinna and Burges, Chris and others},
  url={http://yann. lecun. com/exdb/mnist},
  year={1998}
}

@misc{fashionmnist,
      title={Fashion-MNIST: a Novel Image Dataset for Benchmarking Machine Learning Algorithms}, 
      author={Han Xiao and Kashif Rasul and Roland Vollgraf},
      year={2017},
      eprint={1708.07747},
      archivePrefix={arXiv},
      primaryClass={cs.LG},
      url={https://arxiv.org/abs/1708.07747}, 
}

@inproceedings{yangMedMNISTClassificationDecathlon2021a,
	title = {{MedMNIST} {Classification} {Decathlon}: {A} {Lightweight} {AutoML} {Benchmark} for {Medical} {Image} {Analysis}},
	shorttitle = {{MedMNIST} {Classification} {Decathlon}},
	url = {http://arxiv.org/abs/2010.14925},
	doi = {10.1109/ISBI48211.2021.9434062},
	urldate = {2025-10-30},
	booktitle = {2021 {IEEE} 18th {International} {Symposium} on {Biomedical} {Imaging} ({ISBI})},
	author = {Yang, Jiancheng and Shi, Rui and Ni, Bingbing},
	month = apr,
	year = {2021},
	note = {arXiv:2010.14925 [cs]},
	pages = {191--195},
    eprint={2010.14925},
    archivePrefix={arXiv},
    primaryClass={cs.CV},
}

@article{medmnistv2,
    title={MedMNIST v2-A large-scale lightweight benchmark for 2D and 3D biomedical image classification},
    author={Yang, Jiancheng and Shi, Rui and Wei, Donglai and Liu, Zequan and Zhao, Lin and Ke, Bilian and Pfister, Hanspeter and Ni, Bingbing},
    journal={Scientific Data},
    volume={10},
    number={1},
    pages={41},
    year={2023},
    publisher={Nature Publishing Group UK London},
	url = {http://arxiv.org/abs/2110.14795},
	doi = {10.1038/s41597-022-01721-8},
	eprinttype = {arxiv},
	eprint = {2110.14795},
    primaryClass = {cs, eess},
}

@article{mei2022radimagenet,
	title = {{RadImageNet}: {An} open radiologic deep learning research dataset for effective transfer learning},
	volume = {4},
	doi = {10.1148/ryai.210315},
	number = {5},
	journal = {Radiology: Artificial Intelligence},
	author = {Mei, Xueyan and Liu, Zelong and Robson, Philip M and Marinelli, Brett and Huang, Mingqian and Doshi, Amish and Jacobi, Adam and Cao, Chendi and Link, Katherine E and Yang, Thomas and {others}},
	year = {2022},
	note = {Publisher: Radiological Society of North America},
	pages = {e210315},
}

@misc{krizhevsky2009learning,
  title={Learning multiple layers of features from tiny images.(2009)},
  author={Krizhevsky, Alex and Hinton, Geoffrey and others},
  year={2009},
  url={https://www.cs.toronto.edu/~kriz/learning-features-2009-TR.pdf}
}

@article{linMicrosoftCOCOCommon2015,
      title={Microsoft COCO: Common Objects in Context}, 
      author={Tsung-Yi Lin and Michael Maire and Serge Belongie and Lubomir Bourdev and Ross Girshick and James Hays and Pietro Perona and Deva Ramanan and C. Lawrence Zitnick and Piotr Dollár},
      year={2015},
      eprint={1405.0312},
      archivePrefix={arXiv},
      primaryClass={cs.CV},
      url={https://arxiv.org/abs/1405.0312}, 
}

@article{wuCOVIDxCXR4Expanded2023,
      title={COVIDx CXR-4: An Expanded Multi-Institutional Open-Source Benchmark Dataset for Chest X-ray Image-Based Computer-Aided COVID-19 Diagnostics}, 
      author={Yifan Wu and Hayden Gunraj and Chi-en Amy Tai and Alexander Wong},
      year={2023},
      eprint={2311.17677},
      archivePrefix={arXiv},
      primaryClass={eess.IV},
      url={https://arxiv.org/abs/2311.17677}, 
}

@inproceedings{aggarwal2001surprising,
  title={On the surprising behavior of distance metrics in high dimensional space},
  author={Aggarwal, Charu C and Hinneburg, Alexander and Keim, Daniel A},
  booktitle={International conference on database theory},
  pages={420--434},
  year={2001},
  organization={Springer}
}

@article{aaronson2025,
      title={Limits to black-box amplification in QMA}, 
      author={Scott Aaronson and Phillip Harris and Freek Witteveen},
      year={2025},
      eprint={2509.21131},
      archivePrefix={arXiv},
      primaryClass={quant-ph},
      url={https://arxiv.org/abs/2509.21131}, 
}

@article{guevara2026,
      title={Single-minus gluon tree amplitudes are nonzero}, 
      author={Alfredo Guevara and Alexandru Lupsasca and David Skinner and Andrew Strominger and Kevin Weil},
      year={2026},
      eprint={2602.12176},
      archivePrefix={arXiv},
      primaryClass={hep-th},
      url={https://arxiv.org/abs/2602.12176}, 
}

@article{chang2026,
      title={Exploring Collatz Dynamics with Human-LLM Collaboration}, 
      author={Edward Y. Chang},
      year={2026},
      eprint={2603.11066},
      archivePrefix={arXiv},
      primaryClass={math.DS},
      url={https://arxiv.org/abs/2603.11066}, 
}

@misc{wang2026,
      title={HorizonMath: Measuring AI Progress Toward Mathematical Discovery with Automatic Verification}, 
      author={Erik Y. Wang and Sumeet Motwani and James V. Roggeveen and Eliot Hodges and Dulhan Jayalath and Charles London and Kalyan Ramakrishnan and Flaviu Cipcigan and Philip Torr and Alessandro Abate},
      year={2026},
      eprint={2603.15617},
      archivePrefix={arXiv},
      primaryClass={cs.LG},
      url={https://arxiv.org/abs/2603.15617}, 
}

@article{maxfield1962,
  title={On the matrix equation $X' X = A$},
  author={Maxfield, John E and Minc, Henryk},
  journal={Proceedings of the Edinburgh Mathematical Society},
  volume={13},
  number={2},
  pages={125--129},
  year={1962},
  publisher={Cambridge University Press},
  doi={10.1017/S0013091500014681},
  url={https://www.cambridge.org/core/journals/proceedings-of-the-edinburgh-mathematical-society/article/on-the-matrix-equation-xx-a/37FE330833F83FE01B21DC8CBB85209F}
}

@article{frobenius1912,
  title={{\"U}ber Matrizen aus nicht negativen Elementen},
  author={Frobenius, Ferdinand Georg},
  year={1912},
  publisher={K{\"o}nigliche Akademie der Wissenschaften Berlin, Germany},
  url={https://upload.wikimedia.org/wikipedia/commons/4/44/Ueber_Matrizen_aus_nicht_negativen_Elementen.pdf}
}

@article{perron1907,
	title = {Zur {Theorie} der {Matrices}},
	volume = {64},
	copyright = {http://www.springer.com/tdm},
	issn = {0025-5831, 1432-1807},
	url = {http://link.springer.com/10.1007/BF01449896},
	doi = {10.1007/BF01449896},
	language = {de},
	number = {2},
	urldate = {2026-03-19},
	journal = {Mathematische Annalen},
	author = {Perron, Oskar},
	month = jun,
	year = {1907},
	pages = {248--263},
}

@article{hoffman1966,
    title={Three Observations on Nonnegative Matrices},
    author={Hoffman, A. J.},
    journal={Journal of Research of The National Bureau of Standards-B. Mathematics and Mathematical Physics},
    volume={71B},
    number={1},
    pages={39-41},
    year={1966},
    publisher={NIST},
    doi = {10.6028/jres.071B.007},
	url = {https://nistdigitalarchives.contentdm.oclc.org/digital/collection/p16009coll6/id/87637/rec/1},
}

@article{london1966,
    title={Two inequalities in nonnegative symmetric matrices},
    author={London, David},
    journal={Pacific Journal of Mathematics},
    volume={16},
    number={3},
    pages={515-536},
    year={1966},
    publisher={Mathematical Sciences Publishers},
    doi = {10.2140/pjm.1966.16.515},
	url = {https://msp.org/pjm/1966/16-3/p11.xhtml},
}

@ARTICLE{Araki1990,
       author = {{Araki}, Huzihiro},
        title = "{On an inequality of Lieb and Thirring}",
      journal = {Letters in Mathematical Physics},
         year = 1990,
        month = feb,
       volume = {19},
       number = {2},
        pages = {167-170},
          doi = {10.1007/BF01045887},
       adsurl = {https://ui.adsabs.harvard.edu/abs/1990LMaPh..19..167A}
}

@article{Lieb:1976ny,
    author = "Lieb, Elliott H. and Thirring, Walter E.",
    title = "{Inequalities for the Moments of the Eigenvalues of the Schrodinger Hamiltonian and their Relation to Sobolev Inequalities}",
    reportNumber = "Print-76-0039 (PRINCETON)",
    pages = "205--239",
    month = "1",
    year = "1976"
}

@article{Weyl:1912,
	title = {Das asymptotische {Verteilungsgesetz} der {Eigenwerte} linearer partieller {Differentialgleichungen} (mit einer {Anwendung} auf die {Theorie} der {Hohlraumstrahlung})},
	volume = {71},
	copyright = {http://www.springer.com/tdm},
	issn = {0025-5831, 1432-1807},
	url = {http://link.springer.com/10.1007/BF01456804},
	doi = {10.1007/BF01456804},
	language = {de},
	number = {4},
	urldate = {2026-03-20},
	journal = {Mathematische Annalen},
	author = {Weyl, Hermann},
	month = dec,
	year = {1912},
	pages = {441--479},
}

@misc{audenaert2007,
      title={On the Araki-Lieb-Thirring inequality}, 
      author={Koenraad M. R. Audenaert},
      year={2007},
      eprint={math/0701129},
      archivePrefix={arXiv},
      primaryClass={math.FA},
      url={https://arxiv.org/abs/math/0701129}, 
}

\end{document}